\documentclass[sigconf]{acmart}
\usepackage{amsmath, enumitem, overpic, multirow, xcolor}

\AtBeginDocument{%
  \providecommand\BibTeX{{%
    \normalfont B\kern-0.5em{\scshape i\kern-0.25em b}\kern-0.8em\TeX}}}

\setcopyright{none}
\renewcommand\footnotetextcopyrightpermission[1]{} % removes footnote with conference information in first column
\begin{document}

\title{Drag Your Gaussian: Effective Drag-Based Editing  with Score Distillation for 3D Gaussian Splatting}

\definecolor{color_red}{rgb}{1, 0, 0} % RGB颜色模型
\definecolor{color_blue}{rgb}{0.02, 0, 1} % RGB颜色模型
\definecolor{color_grey}{rgb}{0.58, 0.58, 0.58} % RGB颜色模型

\definecolor{color_orange}{rgb}{0.97, 0.69, 0.5} % RGB颜色模型
\definecolor{color_green}{rgb}{0.62, 0.82, 0.53} % RGB颜色模型

\author{
Yansong Qu$^{1}$, 
Dian Chen$^{1}$, 
Xinyang Li$^{1}$, 
Xiaofan Li$^{\textsuperscript{2\dag}}$,\\
Shengchuan Zhang$^{1}$, 
Liujuan Cao$^{\textsuperscript{1\ddag}}$, 
Rongrong Ji$^{1}$ \\
\vspace{0.5em}
\textbf{\LARGE $^1$Xiamen University, $^2$Baidu Inc.}
\vspace{0.5em}
}
 % 这里可替换为 \medskip 或 \smallskip
\affiliation{%
  % \institution{
  % % Key Laboratory of Multimedia Trusted Perception and Efficient Computing, Ministry of Education of China, \\
  %   $^1$Xiamen University, 
  % % $^2$Tencent Youtu Lab, Shanghai, China, \\
  % % $^2$Institute of Artificial Intelligence, Xiamen University, Fujian, China
  %  $^2$Baidu Inc.
  % }
  %\city{$^2$TAL Education Group}
  \country{
  \textsuperscript{\dag}project lead, 
  \textsuperscript{\ddag}corresponding author
  }
}
\email{{quyans, chendian}@stu.xmu.edu.cn, imlixinyang@gmail.com, shalfunnn@gmail.com}
\email{{zsc_2016, caoliujuan, rrji}@xmu.edu.cn}

% \renewcommand{\thefootnote}{}

% \renewcommand{\thefootnote}{}

\renewcommand{\shortauthors}{author name and author name, et al.}


\begin{abstract}
Recent advancements in 3D scene editing have been propelled by the rapid development of generative models.
Existing methods typically utilize generative models to perform text-guided editing on 3D representations, such as 3D Gaussian Splatting (3DGS). However, these methods are often limited to texture modifications and fail when addressing geometric changes, such as editing a character's head to turn around. Moreover, such methods lack accurate control over the spatial position of editing results, as language struggles to precisely describe the extent of edits. 
To overcome these limitations,  we introduce DYG, an effective 3D drag-based editing method for 3D Gaussian Splatting. It enables users to conveniently specify the desired editing region and the desired dragging direction through the input of 3D masks and pairs of control points, thereby enabling precise control over the extent of editing. 
DYG integrates the strengths of the implicit triplane representation to establish the geometric scaffold of the editing results, effectively overcoming suboptimal editing outcomes caused by the sparsity of 3DGS in the desired editing regions.
Additionally, we incorporate a drag-based Latent Diffusion Model into our method through the proposed Drag-SDS loss function,
enabling flexible, multi-view consistent, and fine-grained editing.
Extensive experiments demonstrate that DYG conducts effective drag-based editing guided by control point prompts, surpassing other baselines in terms of editing effect and quality, both qualitatively and quantitatively.
Visit our project page at \url{https://quyans.github.io/Drag-Your-Gaussian/}.

% \footnotetext{\textsuperscript{†} Corresponding Author. }
\end{abstract}

% <concept_id>10010147.10010371.10010372</concept_id>
% <concept_desc>Computing methodologies~Rendering</concept_desc>

% \ccsdesc[500]{Computing methodologies~Rendering}


\begin{teaserfigure}
    \centering
    \includegraphics[width=\textwidth]{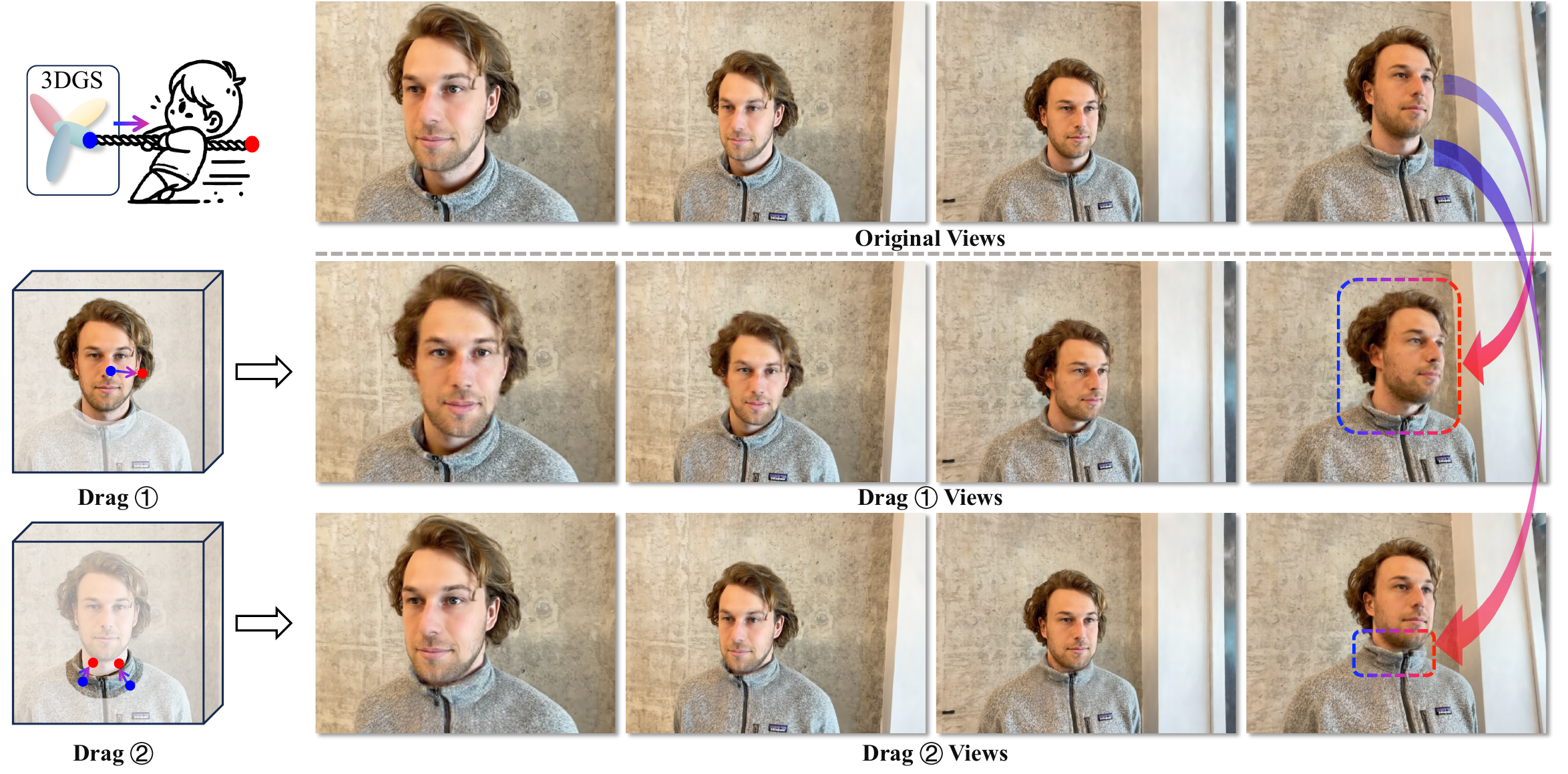}
    \caption{
     % DYG achieves flexible, high quality drag-based 3D editing results. Given a reconstructed 3D Gaussinan field,
  % 用户通过pairs of control points, 包括 handle points （蓝色）、target points（红色），以及3D masks （brighter area） to 指定目标编辑区域
  DYG achieves flexible, high-quality drag-based 3D editing results. Given a reconstructed 3D Gaussian field, users specify the desired editing area with 3D masks (\textcolor{color_grey}{brighter} areas), and perform scene edits 
  through pairs of control points, including handle points (\textcolor{color_blue}{blue}), target points (\textcolor{color_red}{red}).
    }
    \Description{figure description}
    \label{fig-teaser}
  \end{teaserfigure}

\maketitle

\section{Introduction}

The representation and manipulation of 3D scenes have  become increasingly significant in a variety of fields, such as virtual reality (VR) and augmented reality (AR).
Traditional 3D representations, such as meshes, voxels, and point clouds, have facilitated many advancements but typically encounter challenges in scalability, efficiency, and expressiveness. The developments in neural scene representations, such as Neural Radiance Fields (NeRF) \cite{mildenhall2021nerf}, have suggested impressive capabilities in synthesizing photorealistic novel views. However, NeRF-related methods \cite{chen2022tensorf, muller2022instant, wang2023ripnerf, qu2023sgnerf} rely on extensive sampling processes and are computationally intensive, making them less suitable for interactive editing tasks.

\begin{figure}
    \centering
    \includegraphics[width=\linewidth]{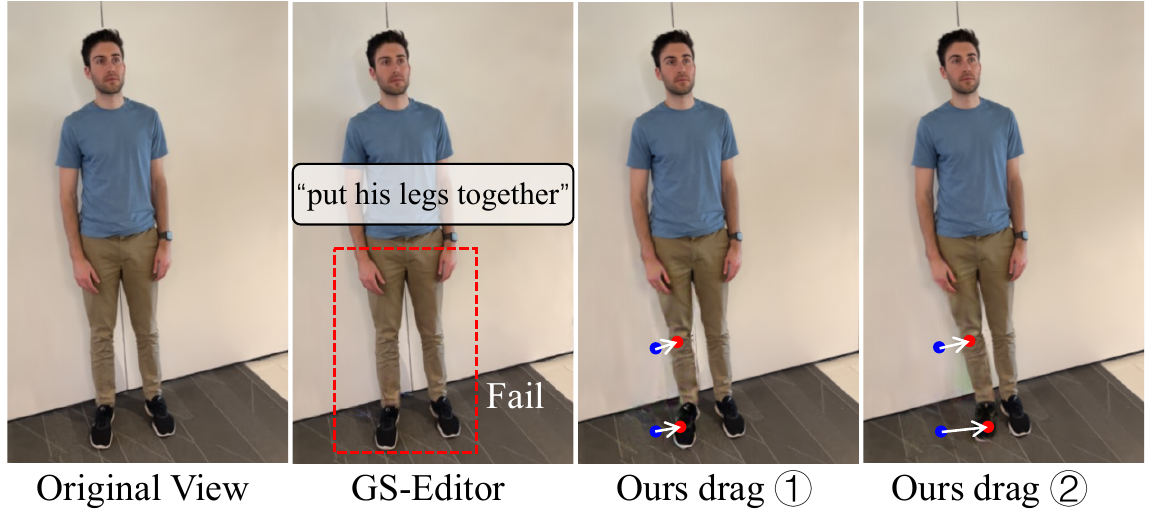}
    \caption{Differences between our drag-based editing approach and the text-guided editing method GS-Editor  \cite{chen2024gaussianeditor}. 
    The latter often fails to achieve geometric editing goals and struggles to describe the degree of editing through text, whereas our method allows for  flexible control over the extent of edits.
    }
    \label{fig-motivation}
\end{figure}

Recently, 3D Gaussian Splatting (3DGS) \cite{kerbl20233dgaussian} has attracted substantial attention for its ability to represent volumetric data using sparse Gaussian primitives. By replacing dense neural networks with lightweight and interpretable Gaussian primitives, 3DGS enables real-time rendering and fast updates, making it a promising 3D representation for 3D  editing.  
Recent 3DGS-based scene editing methods \cite{chen2024gaussianeditor,chen2025dge,wu2025gaussctrl,wang2024gaussianeditor_water} leverage pre-trained 2D latent diffusion models (LDM) with text prompts to guide the optimization of 3DGS. However, they primarily focus on texture modifications or stylistic changes, falling short in enabling precise geometric editing, as illustrated in Fig. \ref{fig-motivation}. 

To overcome these geometric editing limitations, 
we draw inspiration from recent advancements in 2D drag-based image editing. 
Methods such as DragGAN \cite{pan2023draggan} and its successors \cite{shi2024dragdiffusion, shi2024instadrag} provide  precise control and intuitive image editing capabilities through the use of paired control points, including handle points and target points. 
However, applying 2D drag-based generative models to guide the optimization of 3DGS for drag-based 3D editing introduces a new challenge: the target regions often exhibit sparse distributions of 3D Gaussians, making it challenging to effectively edit the 3D Gaussian field. 
Consequently, the model tends to align the texture of nearby 3D Gaussians around the target area, rather than accurately generating the desired geometric structures. This issue significantly impacts the precision and realism of the editing results.

A straightforward approach involves adopting a rigid transformation \cite{wang2023ripnerf, xu2022deforming}, where 3DGS primitives around the handle points are copied to the target region during initialization. However, it significantly limits the diversity of editing tasks and often results in poor geometry in the target area, leading to noticeable artifacts.

In this work, we present \textbf{D}rag \textbf{Y}our 3D \textbf{G}aussian (DYG), a novel drag-based 3DGS editing approach for real-world scenes. 
To enhance usability and accessibility, we extend the 2D drag-based image editing paradigm to 3D, introducing 3D masks along with pairs of 3D control points as inputs for editing 3D scenes. 
We integrate the independence and discreteness of 3DGS with the continuous nature of implicit triplane representation to address the challenge of sparse 3D Gaussian distributions. To encode the positions of 3D Gaussians, we introduce the Multi-scale Triplane Positional (MTP) Encoder, and employ a Region-Specific Positional (RSP) Decoder to predict positional offsets, constructing the geometric scaffold for dragging. Additionally, we propose a Soft Local Edit (SLE) strategy to focus editing on the desired region while preserving the integrity of other areas. Leveraging an off-the-shelf 2D drag-based LDM as supervision through the proposed Drag-SDS loss function, we enable perceptually plausible scene dragging with multi-view consistency. As shown in Fig. \ref{fig-teaser} and Fig. \ref{fig-motivation}, DYG facilitates flexible and fine-grained 3D scene editing.

Our contributions can be summarized as follows:

\begin{itemize}[leftmargin=0.5cm]

\item   We propose an effective drag-based scene editing method for 3D Gaussian Splatting,  capable of delivering flexible and high-quality results for geometric editing tasks, including deformation, transformation, and morphing.

\item 
 We introduce the MTP encoder to address the challenge of uneven spatial distribution of Gaussian primitives, facilitating smooth geometric editing. Additionally, the RSP decoder and SLE strategy ensure harmonious local editing. Finally, Drag-SDS leverages the existing 2D drag-based LDM to achieve multi-view consistent dragging results.

 \item Extensive experiments quantitatively and qualitatively demonstrate that our method achieves state-of-the-art (SOTA) 3D scene editing results, validating the versatility and generalization capabilities of DYG.

\end{itemize}

\begin{figure*}[ht]
    \centering
    \includegraphics[width=\linewidth]{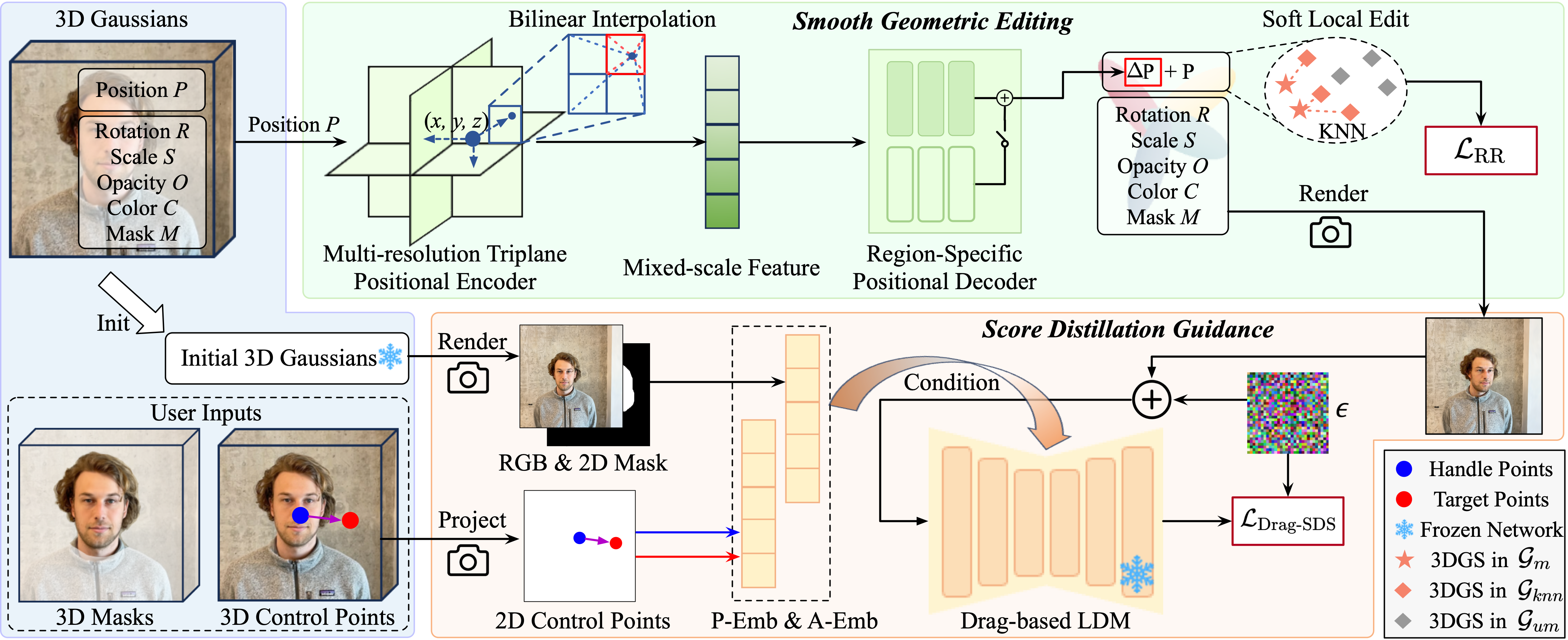}
    \caption{
    The overall framework of DYG. 
    \textbf{Left:} 
    Given a 3D Gaussian scene, users provide 3D masks and several pairs of control points as input. 
    \textbf{Top-right:} The Smooth Geometric Editing module predicts positional offsets for 3D Gaussians, resolving the issue of sparse distributions within the target region while ensuring seamless local editing. We adopt a two-stage training strategy: the first stage constructs the geometric scaffold of the edited Gaussians, and the second stage refines the texture details.
    \textbf{Bottom-right:} 
    In the Score Distillation Guidance Module, to ensure stable optimization, 3D control points are projected onto 2D control points for a specified viewpoint. The RGB image and 2D mask, rendered from the mirrored initial 3D Gaussians, are encoded into point embeddings (P-Emb) and appearance embeddings (A-Emb), which act as conditions for the drag-based LDM. This process leverages our proposed Drag-SDS loss function to enable flexible and view-consistent 3D drag-based editing.
    }
    \label{fig-pipeline}
\end{figure*}

\section{Related Work}
\subsection{{Drag-based Image Editing}}

In light of advancements in generative models, image editing has seen significant development. However, text-guided image editing methods \cite{ramesh2022hierarchical, brooks2023instructpix2pix, hertz2022prompt, kawar2023imagic} often lack precision and flexibility when it comes to editing spatial attributes.
To address this, DragGAN \cite{pan2023draggan} enables impressive interactive drag-based image editing by utilizing control points and optimizing generative adversarial networks (GANS) latent codes.
However, the applicability of this framework is constrained by the intrinsic limitations in the capacity of GANs. 
In order to enhance generalization,
subsequent works \cite{luo2024readout, shi2024dragdiffusion, mou2023dragondiffusion} extend this paradigm to large-scale Latent Diffusion Models (LDM). 
However, these methods depend on computationally intensive operations, such as latent optimization, resulting in inefficiencies in editing tasks.

Lightning-Drag \cite{shi2024instadrag} encodes user prompts into corresponding point embeddings, which are then injected into the self-attention modules of the Stable Diffusion inpainting backbone \cite{ho2020ddpm, song2020ddim} to guide the generation process. This approach eliminates the need for time-consuming operations required in previous methods \cite{nie2023blessing, shi2024dragdiffusion}, enabling interactive drag-based image editing.
In this work, we adopt Lightning-Drag \cite{shi2024instadrag} as the guiding model for editing 3DGS, owing to its rapid and high-quality drag-based editing capabilities.

\subsection{3D Editing for Radiance Fields}
Neural Radiance Fields (NeRF) \cite{mildenhall2021nerf} introduced radiance fields,
 having excelled in novel view synthesis, producing 
realistic rendering results. However, NeRF’s reliance on a neural
network for complete implicit representation of scenes leads to
tedious training and rendering times. More recently, 3DGS [9] has garnered
attention from researchers due to its real-time rendering speed
and photo-realistic rendering quality.

Robust 3D representations have driven advancements in 3D editing. Early methods \cite{guo2020object, yang2021learning, ost2021neural} learn object-compositional NeRFs, enabling object-level editing, such as duplicating or moving objects. However, these approaches are limited to coarse-grained manipulations.
Works like NeuMesh \cite{yang2022neumesh} and related methods \cite{xu2022deforming, yuan2022nerf, wang2023ripnerf} propose using explicit geometry, such as cages or point clouds, to facilitate geometric editing. Nevertheless, these methods heavily rely on precise geometric reconstructions. SC-GS \cite{huang2024scgs} adopts a sampling-based approach to learn anchor points for editing 3D scenes. However, these methods strongly depend on accurate geometric representations, offer limited editing diversity, and often suffer from unreasonable results, such as local tearing artifacts.

With the success of generative models, methods \cite{haque2023instruct, zhuang2023dreameditor, zhuang2024tip, wang2024gaussianeditor_water},  like GS-Editor \cite{chen2024gaussianeditor} and GS-Ctrl \cite{wu2025gaussctrl} leverage text-guided latent diffusion models for 3D scene editing. While effective for texture or style modifications, these approaches often fail to handle geometric changes. Moreover, they struggle to accurately specify the spatial extent of editing through text.
In contrast, our method enables flexible, fine-grained geometric editing while maintaining the plausibility of the edited scenes.

\section{Preliminary}
\label{subsec_3dgs}

\subsection{3D Gaussian Splatting}  
3D Gaussian Splatting (3DGS) \cite{kerbl20233dgaussian}  utilizes a set of anisotropic 3D Gaussians to model three-dimensional information and provide fast rendering by efficiently rasterizing 3D Gaussians into images, given camera poses. 
Specifically, each Gaussian is composed of its position $p \in \mathbb R^3$, 
a scale $s \in \mathbb R^3$, a rotation quaternion $r \in \mathbb R^4$,
an opacity $o \in \mathbb R $, and the spherical harmonics (SH) coefficients $c \in \mathbb R^d$ for volume rendering. These 3D Gaussians are projected onto the image plane as 2D Gaussians and rendered in real time using the tiled rasterizer.

\subsection{Score Distillation Sampling (SDS)} 
DreamFusion \cite{poole2022dreamfusion}  introduces the SDS loss function,  utilizing a pre-trained 2D latent diffusion model (LDM) to optimize 3D representations for 3D generation.
Specifically, given a differentiable 3D representation, such as NeRF \cite{mildenhall2021nerf} or 3DGS \cite{kerbl20233dgaussian}, parameterized by  $\mathcal{G}$  and a rendering function  $\mathcal{R}$, the rendered image corresponding to a camera pose  $c$  can be expressed as  $x = \mathcal{R}(\mathcal{G}, c)$. SDS leverages the prior knowledge of a LDM to guide the optimization of the 3D representation  $\mathcal{G}$  
in a low-resolution latent space. This latent space is articulated as $z = \mathcal{E}(x), x = \mathcal{D}(z)$, where $\mathcal{E}$ and  $\mathcal{D}$ represent the encoder and decoder of the LDM, respectively. 
% LDM 
The SDS loss function is formulated as follows:
\begin{equation}
    \nabla_{\mathcal{G}} \mathcal{L}_{\mathrm{SDS}}=\mathbb{E}_{t, \epsilon, c}\left[w(t)(\hat{\epsilon}-\epsilon) \frac{\partial \mathcal{E}(\mathcal{R}(\mathcal{G}, c))}{\partial \mathcal{\mathcal{G}}}\right]
    \label{eq-sds}
\end{equation}
where $\epsilon$ denotes ground truth noise,  $\hat{\epsilon}$ is the noise predicted by the LDM with  $z_t$  as input for timestep  $t$, and $w(t)$ represents a weighting function that varies according to the timestep $t$. 
The SDS loss can be reformulated  
\cite{song2020ddim, zhu2023hifa} as follows:
\begin{equation}
    \mathcal{L}_{\mathrm{SDS}}=\mathbb{E}_{t, \epsilon, c}\left[w(t) \frac{\sqrt{\bar{\alpha}_t}}{\sqrt{1-\bar{\alpha}_t}}\|z-\hat{z}\|_2^2\right],
    \label{eq-sds-recon}
\end{equation}
where 
\begin{equation}
    \hat{z} = \frac{z_t - \sqrt{1-\bar{\alpha}_t}\hat{\epsilon}}
    {\sqrt[]{\bar{\alpha}_t}},
\end{equation}
with $\bar{\alpha}(t)$ is also a weighting function that dynamically varies with each timestep $t$.

\section{Methods}
\subsection{Problem Definition and Method Overview}
\label{subsec_overview}

Given a reconstructed 3D Gaussian field  $\mathcal{G}$, we extend the attributes of 3D Gaussians by adding a mask attribute $m$, which represents the user-defined 3D masks for the desired editing area. The Gaussian field  $\mathcal{G}$ can be defined as:
$ \mathcal{G} =\left\{{p}_{i}, {r}_{i}, {s}_{i}, \alpha_{i}, {c}_{i},{m}_{i}\right\}_{i=1}^{N}$.
In addition to the 3D mask, users are also required to input $K$ pairs of control points 
$Q=\{(q_i^o ,q_i^t) \}_{i=1}^{K}$, where $q_i^o ,q_i^t \in \mathbb R^3$ serve as guidance for the editing process. 
Our objective is to drag the the desired editing region around the handle points $q_i^o$ to the target points $q_i^t$.

\textit{\textbf{Rendering and Projection for 2D Guidance.} }
In order to ensure stable control of the 2D drag-based LDM, a mirrored copy of  $\mathcal{G}$  is preserved,  referred to as  the Initial 3D Gaussians  $\mathcal{G'}$, as shown in Fig. \ref{fig-pipeline}.
During the training phase, for a given camera pose  $c$, an RGB image  $I_c$  and a 2D mask are rendered from  $\mathcal{G'}$  using the similar volumetric rendering approach \cite{qu2024goi,kerbl20233dgaussian}.
This 2D mask is then subjected to a dilation operation to produce the final 2D mask $M_c$. Additionally, the 3D control points $Q$ are projected into 2D points $Q_{c}^{2d} = \{ (\Pi(\mathbf{q}_i^o), \Pi(\mathbf{q}_i^t)) \}_{i=1}^{K},$ through the projection transformation $\Pi$. 
Upon acquiring $I_c$, $M_c$ and view-specific 2D control points $Q_c^{2d}$ obtained, these inputs are utilized as the condition $y$ of
2D drag-based LDM to guide 3DGS optimization using our proposed Drag-SDS loss.

\textit{\textbf{Local Edit.} } 
Our guiding principle is to perform localized edits within the desired editing region while ensuring that the rest of the scene remains unaffected, thereby maintaining overall harmony and realism.

To facilitate this process, we have developed an interactive GUI tool based on 3D Gaussian Splatting. This tool enables users to identify desired editing regions from different viewpoints and generates the 3D mask by calculating the intersection of 3D Gaussians within each view frustum. This real-time interactive approach allows users to efficiently complete the 3D mask selection process with minimal effort,  potentially requiring only a single operation.

\textit{\textbf{Method Overview.} } 
Figure~\ref{fig-pipeline} illustrates the  overview of our method. 
In Sec. \ref{subsec_Smooth_Geometric_Editing}, we explain how to integrate the strengths of the implicit triplane representation and explicit 3DGS to overcome suboptimal editing outcomes caused by the sparsity of 3DGS in target regions, thereby enabling high-quality, localized drag-based editing.
In Sec. \ref{subsec_Score_distillation_guidance}, we describe how the existing 2D drag-based LDM is incorporated into our method through the proposed Drag-SDS loss function, enabling flexible, view-consistent, and fine-grained editing.

% \vspace{-5mm}

\subsection{Smooth Geometric Editing}
\label{subsec_Smooth_Geometric_Editing}

3D drag-based editing encompasses three main scenarios:
(1) Deformation: Involves fine-grained edits, such as adjusting facial features to face a different direction. 
(2) Transformation: Encompasses local rigid transformations, exemplified by moving a man’s leg to take a step forward. 
(3) Morphing: Includes structural adjustments, such as to raise the collar or to make a person’s shoulders narrower. 

For scenario (1), the challenge lies in fine-grained local editing—modifying the desired region while preserving other areas as much as possible.
For scenarios (2) and (3), the key challenge lies in the sparsity of 3D Gaussians around the target points, making it difficult to generate new Gaussians within the target region through optimization or the densify and prune operations \cite{kerbl20233dgaussian}. This often results in editing failures.

Therefore, based on the above observations and to propose a unified solution, we present the Multi-resolution Triplane Positional Encoder, Region-Specific Positional Decoder, Two-stage Dragging, and Soft Local Editing strategies to achieve smooth geometric editing. 

\textit{\textbf{Multi-resolution Triplane Positional (MTP) Encoder.} } 
The triplane representation \cite{chan2022eg3d,zou2024triplane,fridovich2023kplane} is distinguished by its compactness and efficient expressiveness.
Its implicit nature  facilitating the learning of 3D structures through volume rendering \cite{qu2023sgnerf,kerbl20233dgaussian}, providing an  effective solution to the uneven spatial distribution of Gaussian primitives.

Another consideration is that the intuitive approach for 3D drag-based editing should involve moving the original region’s Gaussians primitives to the target region, rather than deleting the primitives in the original region and generating new ones in the target region. To achieve this, we introduce the Multi-resolution Triplane Positional (MTP) Encoder to encode the position of the 3D Gaussians and predict the position shifts $\Delta P$ with the Region-Specific Positional Decoder.

Specifically, the MTP decomposes the 3D space into three orthogonal, learnable multi-resolution feature planes: $\mathcal{H}{xy}, \mathcal{H}{xz}, \mathcal{H}_{yz}.$ 
For the position $p = (x, y, z)$ of each 3D Gaussian, it is normalized and projected onto these triplanes at varying resolutions:
\begin{equation}
    f_c^s = \psi^s \left ( \mathcal{H}_c, \pi_c \left(x,y,z \right) \right ),
\end{equation}
\begin{equation}
    f=\Theta \big(\! \operatorname*{concat}_{s} \prod_{c\in C} f_c^s \big),
\end{equation}
where $\pi_c$ projects the point onto the $c$-th plane, and $\psi^s$ performs bilinear interpolation on the plane at resolution $s$. $\prod$ denotes the Hadamard product, C represents the set of planes $\{xy, xz, yz\}$, and $\Theta$ is a lightweight MLP for fusing positional mixed-scale  features.

% \paragraph{\textbf{Region-Specific Positional Decoders (RSP)}}
\textit{\textbf{Region-Specific Positional  (RSP) Decoder.}}  
We introduce the local editing guiding principle in Sec. \ref{subsec_overview}, however, the implicit representation of triplane, combined with the latent-based optimization of SDS, inevitably result in changes to regions outside the 3D mask. 
To address this issue, the RSP decoder predict the position shifts  $\Delta P$  of the masked 3DGS while introducing a new network to correct the unintended movements in regions outside the 3D masks. 
Additionally, we propose a regularization loss to further constrain the optimization process.
Specifically, 
The 3DGS  $\mathcal{G}$ is divided into two subsets: $\mathcal{G}_{m}= \{ ({p}_{i}, {r}_{i}, {s}_{i}, o_{i}, {c}_{i},{m}_{i} )\in \mathcal{G} \mid m_i=1 \}$ and $\mathcal{G}_{um}= \{ ({p}_{i}, {r}_{i}, {s}_{i}, o_{i}, {c}_{i},{m}_{i} )\in \mathcal{G} \mid m_i=0 \}$.
two MLPs,  $\mathcal{N}_1$ and $\mathcal{N}_2$, are employed as decoders for the feature $f$ obtained from the MTP encoder. The position shift of each Gaussian primitive $g_i$ can be formulated as follows:
\begin{equation}
    \Delta p = 
\begin{cases}
\mathcal{N}_1(f) & \text{for } g_i \in \mathcal{G}_{m} \\
\text{sg}(\mathcal{N}_1(f)) + \mathcal{N}_2(\text{sg}(f)) & \text{for } g_i \in \mathcal{G}_{um},
\end{cases}
\label{RSP-decoder-eq}
\end{equation}
where sg(·) is the stop gradient operator. 
Based on this, we design a region regularization loss to encourage the unmasked Gaussians to remain unchanged:
\begin{equation}
    % \mathcal{L}_\text{RR} = \left\| \Delta P \odot (\vmathbb{1} - M) \right\|_1 ,
    \mathcal{L}_\text{RR} = \sum_{g_i \in \mathcal{G}_{um}} \!\!\! \Delta p_{i}.
\end{equation}

\begin{figure}
    \centering
    \includegraphics[width=\linewidth]{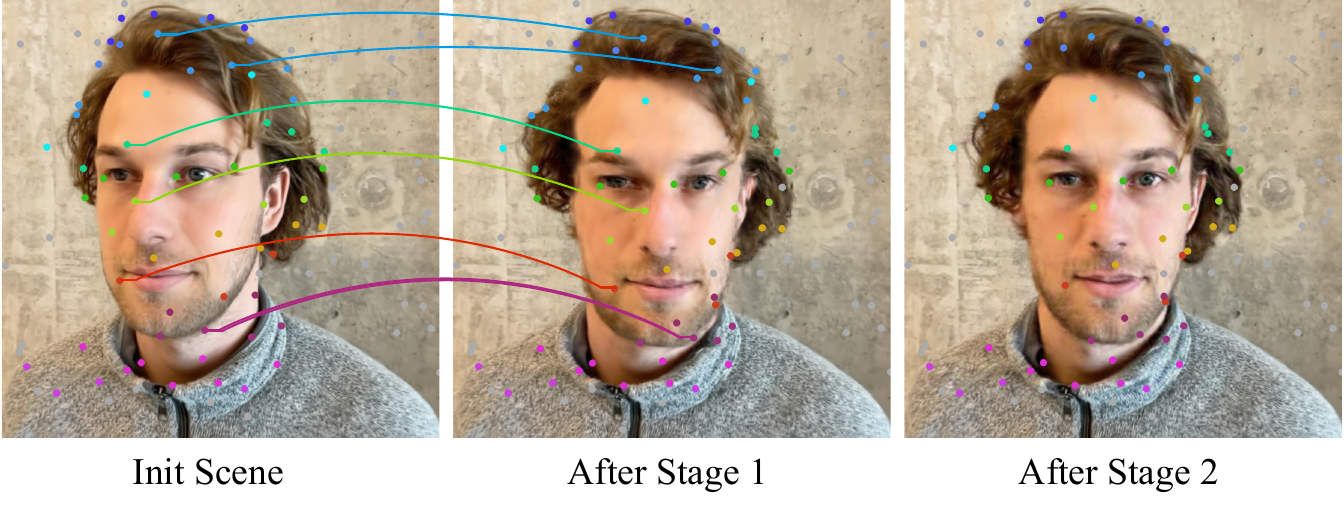}
    \caption{
    % 我们对3D Gaussians进行采样并追踪，可视化 出3D Gaussians在编辑前后的位置变化，他们的位置用点表示。
    % We sample and track 3D Gaussians, visualizing the positional changes of 3D Gaussians before and after editing. Their positions are represented as colored points.
    Visualization of the positional changes of sampled 3D Gaussians after two-stage dragging. The positions of maksed 3D Gaussians are represented as colored points, while others are shown in gray.
    }
    \label{fig-drag-visual}
\end{figure}

\textit{\textbf{Two-stage Dragging.}}
The entire dragging process consists of two stages. The first stage focuses on optimizing the geometric structure, thereby establishing the foundational geometric scaffold of the edited scene. During this stage, the model optimizes both the MTP encoder and RSP decoder while freezing all 3DGS parameters  and halting the densify and prune operations \cite{kerbl20233dgaussian}.  As shown in Fig. \ref{fig-drag-visual} 
In the complex deformation task of turning a character’s face, after Stage 1, the Gaussians corresponding to the face are dragged to the target region, while the positions of unmasked Gaussians remain unchanged.
The second stage focuses on refining the texture details of the scene. During this stage, other attributes of the 3D Gaussians (color, opacity, rotation, and scale) are primarily optimized, and the densify and prune operations are reactivated. 
After constructing the basic geometric scaffold in the first stage, the regions outside the desired editing area tend to remain unchanged during the second stage. The newly added 3D Gaussians primarily originate from the splitting or duplication of elements in  $\mathcal{G}_m$. Consequently, all newly added 3D Gaussians are assigned to  $\mathcal{G}_m$ for subsequent optimization.

\textit{\textbf{Soft Local Edit (SLE).}} 
Strictly freezing the parameters of 3DGS outside the 3D mask \cite{chen2024gaussianeditor,wang2024gaussianeditor_water} to facilitate local editing may result in disjoint effects, as illustrated in Fig. \ref{fig-ablation}(e). To address this limitation, we adopt a soft 3D mask strategy. Specifically, for each 3D Gaussian within $\mathcal{G}{m}$, we identify its K-nearest neighbors (KNN) and select those belonging to $\mathcal{G}{um}$ to form the set $\mathcal{G_{\text{knn}}}$. These neighboring Gaussians are subsequently optimized with a reduced learning rate to ensure smoother transitions.

\subsection{Score Distillation Guidance}
\label{subsec_Score_distillation_guidance}
\begin{figure}
    \centering
    \includegraphics[width=\linewidth]{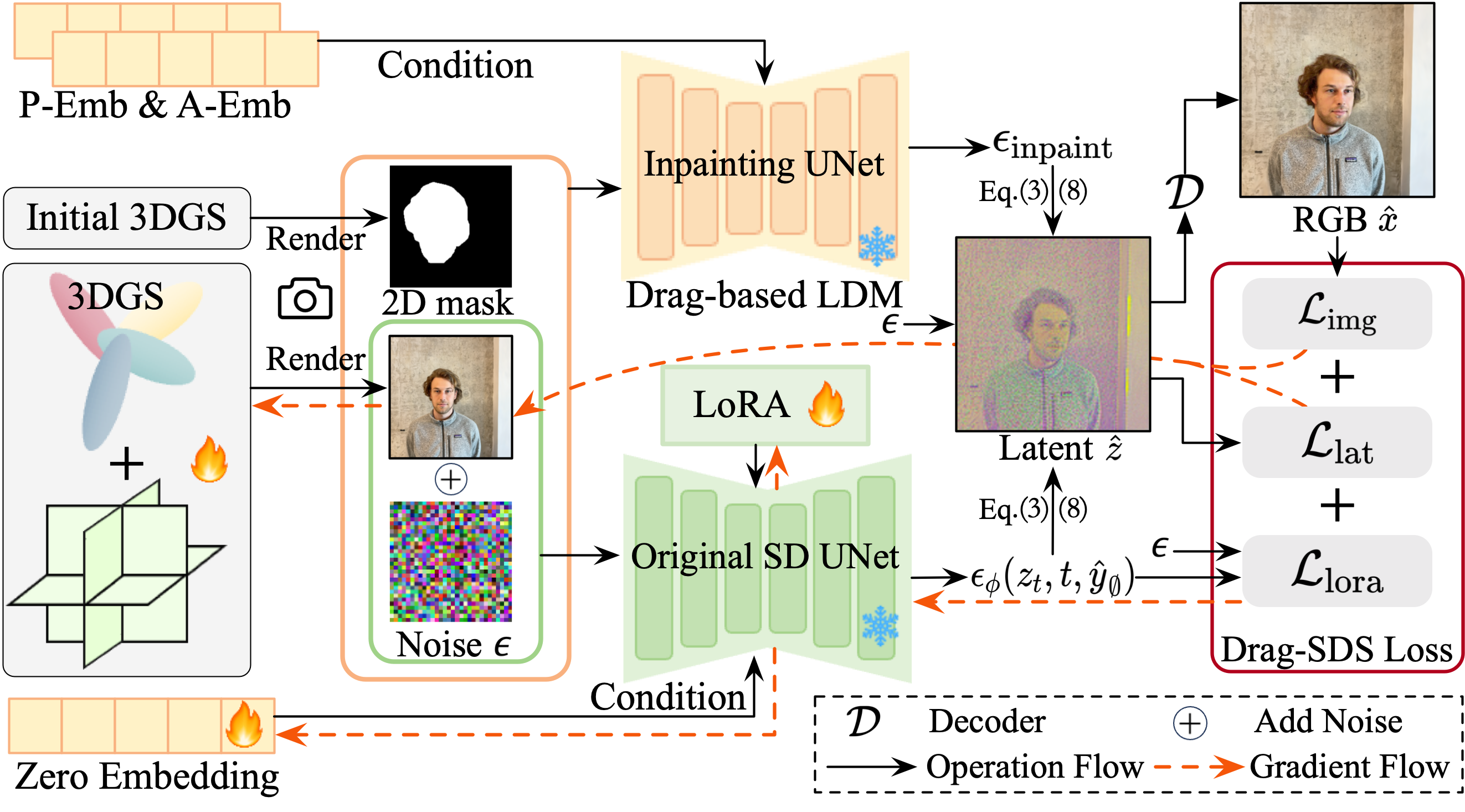}
    \caption{Detailed illustration of Score Distillation Guidance Module and Drag-SDS loss, presented in Fig. \ref{fig-pipeline}.
    We employ two different UNets to predict $\epsilon_{\text{tgt}}$ and $\epsilon_{\text{src}}$ described in Eq. (\ref{eq-composit_epsilon}), respectively. The components within the \textcolor{color_orange}{orange} box represent the inputs to the Inpainting UNet, while the components within the \textcolor{color_green}{green} box signify the inputs to the Original SD UNet. 
    } 
    \label{fig-dsds}
\end{figure}
The SDS loss \cite{poole2022dreamfusion}, which leverages LDM as the guidance model to produce multi-view consistent 3D results, has been widely adopted in 3D generation methods \cite{wang2024prolificdreamer, liu2023zero, zhuang2023dreameditor}. However, it suffers from issues of over-saturation, over-smoothing. 
Inspired by \cite{yang2023lods, li2024director3d}, we extend SDS and propose an improved score distillation loss function. 
For the predicted noise  $\hat{\epsilon}$  in Eq. (\ref{eq-sds}), we extend it into a composite term defined as follows:
\begin{equation}
    \hat{\epsilon} = \epsilon_{\text{tgt}} - \epsilon_{\text{src}} + \epsilon,
    \label{eq-composit_epsilon}
\end{equation}
where $\epsilon_{\text{tgt}}$ represents the noise predicted by the LDM, $\epsilon_{\text{src}}$ denotes a learnable source prediction for adaptive estimation of the current distribution.

Our guidance model, Lightning-Drag \cite{shi2024instadrag}, employs the Stable Diffusion Inpainting U-Net as the backbone to predict $\epsilon_{\text{tgt}}$.  
It takes as input the concatenation of noise latents $z_t$, a binary mask $m_{\text{2d}}$, and the latents of the masked initial image $m_{\text{2d}} \odot x_0$.
 The model also incorporates the point embedding of 2D control points and the appearance embedding as the condition $y$. For clarity, 
we omit the  classifier-free guidance \cite{poole2022dreamfusion, yang2023lods}.  As shown in Fig. \ref{fig-dsds},  the output of our guidance model is represented as 
$\hat{\epsilon}_{\text{inpaint}} =  \epsilon_\theta(z_t, t, y, m_{\text{2d}} , \mathcal{E}(m_{\text{2d}} \odot x_0) )$. Our Drag-SDS loss is composed of three components: image-space loss $\mathcal{L}_\text{img}$, latent-space loss $\mathcal{L}_\text{lat}$, and $\mathcal{L}_\text{lora}$.

\begin{figure*}[ht]
    \centering
    \includegraphics[width=\linewidth]{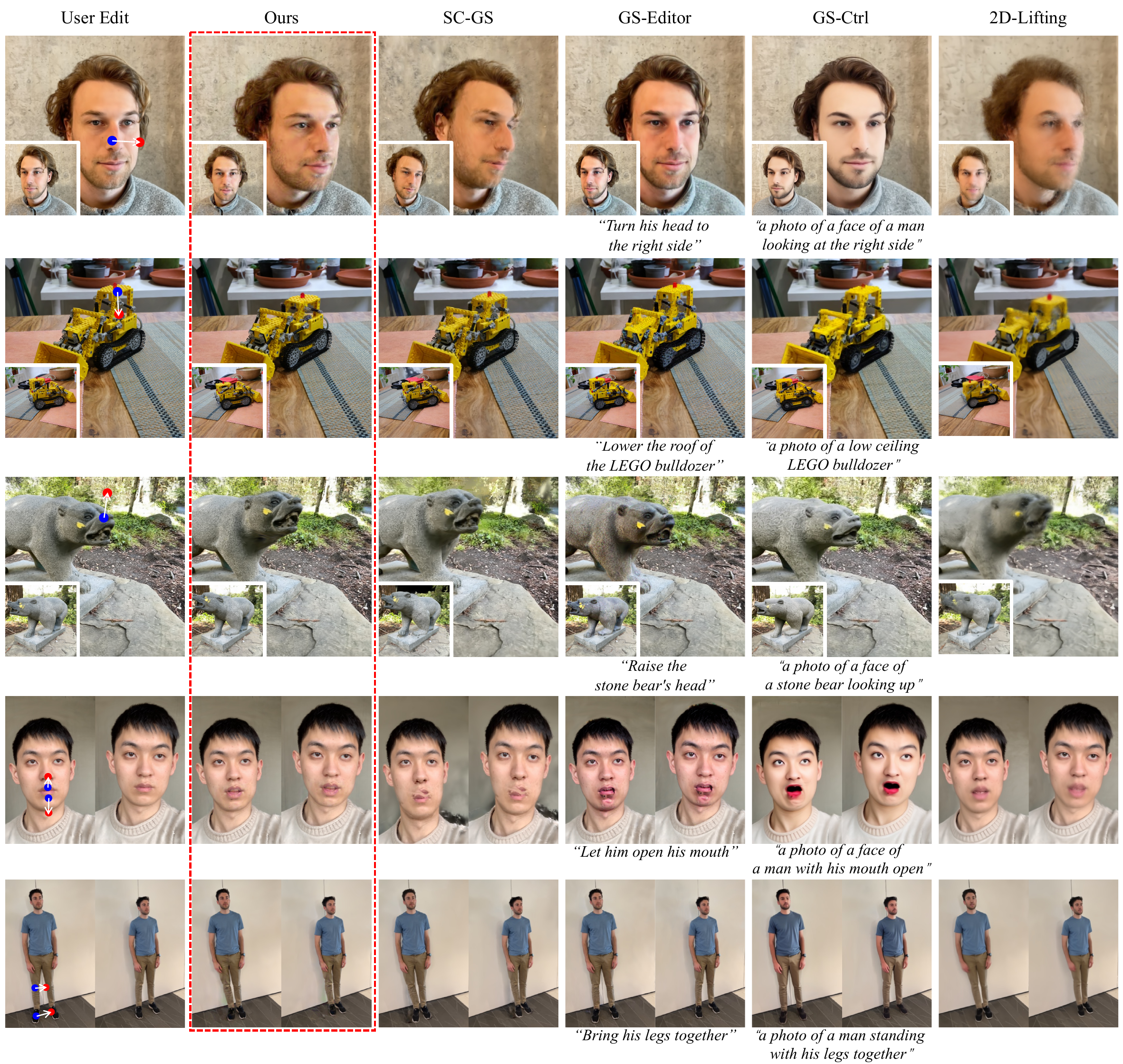}
    \caption{
    % Caption
     Qualitative comparison between DYG and different baselines. 
     % SC-GS 在会出现场景撕裂的现象，GS-Editor and GS-Ctrl会经常会编辑失败，GS-Ctrl还会出现场景过饱和的问题，2D-lifting则会出现场景模糊的现象。相比之下
     % 我们的方法可以实现精细化的编辑并且生成高质量细节的结果。实验展示了我们的方法基于control points能够充分理解用户的意图，从而实现
     The first column shows two rendered views of the original 3D scene, where the 3D editing points are projected onto the 2D plane for visualization. 
      SC-GS \cite{huang2024scgs} may show unnatural results, as well as blurring or tearing of the background, while GS-Editor \cite{chen2024gaussianeditor} and GS-Ctrl \cite{wu2025gaussctrl} frequently fail to perform successful edits. Additionally, GS-Ctrl tends to exhibit over-saturation issues, and 2D-Lifting suffers from scene blurriness. By contrast, 
      DYG is able to sufficiently interpret both the user’s dragging intent and the 3D scene context, thereby achieving effective editing and generating  detailed results across various scenarios, including deformation, transformation, morphing.    }
    \label{fig-comparison}
\end{figure*}

\begin{figure}
    \centering
    \includegraphics[width=\linewidth]{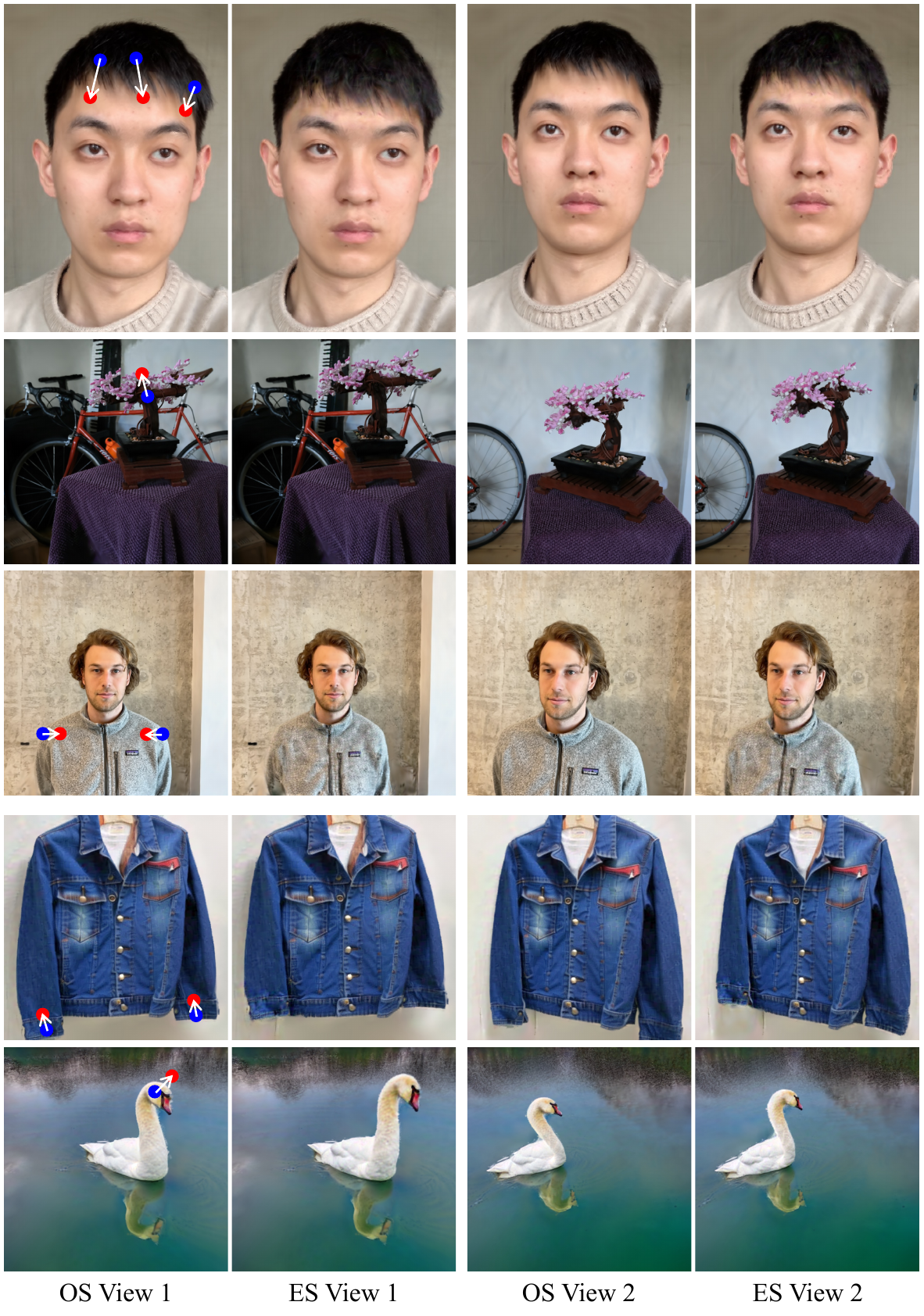}
    \caption{
    % 我们对3D Gaussians进行采样并追踪，可视化 出3D Gaussians在编辑前后的位置变化，他们的位置用点表示。
    % We sample and track 3D Gaussians, visualizing the positional changes of 3D Gaussians before and after editing. Their positions are represented as colored points.
    More qualitative results. The top three rows showcase real scenes, while the bottom two rows are generated scenes.
    For each edit, we show two views of both the original (OS) and edited scenes (ES).
    % 上面三个是真实场景，下面两排是生成场景
    }
    \label{fig-moreresult}
\end{figure}

Lightning-Drag provides reliable predictions for  $\epsilon_{\text{tgt}}$; however, it is not suitable for estimating  $\epsilon_{\text{src}}$. 
This limitation stems from the inpainting backbone tends to focus primarily on the information within the masked region while preserving the content outside the mask, thereby failing to fully capture the current distribution. 
Therefore, different from straightforwardly using the same UNet to predict $\epsilon_{\text{tgt}}$ and $\epsilon_{\text{src}}$ in \cite{yang2023lods, li2024director3d}, we utilize the original Stable Diffusion UNet with a LoRA model \cite{hu2021lora} $\phi$ as the predictor of $\epsilon_{\text{src}}$, denoted as $\hat{\epsilon}_\phi \left(x_t,t,\hat{y}_\emptyset \right)$, where $\hat{y}_\emptyset$ is a learnable embedding initialized to zero. 
The LoRA model is trained using a simple diffusion loss, defined as:
\begin{equation}
\mathcal{L}_{\text{lora}} = \mathbb{E}_{t, c, \epsilon}\left[\| \epsilon_\phi \left(x_t,t,\hat{y}_\emptyset \right) - \epsilon \|_2^2 \right]
\end{equation}
The latent-space score objective, referred to as  $\mathcal{L}_{\text{lat}}$, is formulated similarly to Eq. (\ref{eq-sds-recon}). 
The image-space score distillation loss function is defined as follows: 
\begin{equation}
    \mathcal{L}_{\text{img}} =\mathbb{E}_{t, c, \epsilon}\left[w(t) \frac{\sqrt{\bar{\alpha}_t}}{\sqrt{1-{\bar{\alpha}}_t}}\|x-\hat{x}\|_2^2\right] ,
\end{equation}
% 还差lora loss
where $\hat{x} = \mathcal{D}(\hat{z})$, with $\mathcal{D}$ is the image decoder of LDM.
% 最后我们的分数蒸馏Loss可以综合表示为
The final Drag-SDS loss function can be defined as:
\begin{equation}
    \mathcal{L}_{\text{Drag-SDS}} = \lambda_{\text{lat}}\mathcal{L}_{\text{lat}} + \lambda_{\text{img}}\mathcal{L}_{\text{img}} + \lambda_{\text{lora}}\mathcal{L}_{\text{lora}},
\end{equation}
where $\lambda_{\text{lat}}$, $\lambda_{\text{img}}$, $\lambda_{\text{lora}}$ represent the weights for the  latent-space, image-space, and lora  objectives, respectively.

\section{Implementation Details}

Our model, built on a 3D Gaussian Scene reconstructed using vanilla 3D Gaussian Splatting \cite{kerbl20233dgaussian}, completes one dragging operation on an A100-40G GPU in approximately 10 minutes, whereas GS-Editor \cite{chen2024gaussianeditor} requires over 20 minutes.

Additional implementation details are provided in the Appendix.

\section{Experiments}

\subsection{Evaluation Setup}
% 由于
% 我们在多个场景上进行了实验，用于
% For extensive comparisons,
% 由于我们是第一篇在drag-based real-scene editing的工作，因此我们与其他基于文本编辑高斯的方法，GS-editor，GS-ctrl，进行对比。另外我们构建了一个naive的baseline，名为2D-lifting，使用与我们同样的输入，在2D上使用Lightning-drag进行编辑，然后进行三维重建。

% We distribute 50 copies of questionnaires, presenting rotation video results of all methods side by side and asking users to choose the best editing result.

\paragraph{\textbf{{Baseline Methods.} }}
% 使用了Clip美学指标， 用户调查，以及GPT score
% As the first work on drag-based real-scene editing, 
% It should be noted that this is the first work to perform
% 3DGS drag-based editing in real scenes, and therefore there are no directly comparable baselines.
% we compare our method with other 3DGS-based text-driven 3D  editing approaches, including GS-Editor \cite{chen2024gaussianeditor} and GS-Ctrl \cite{wu2025gaussctrl}. Additionally, the anchor point-based object dragging approach, SC-GS \cite{huang2024scgs}, is included for comparison.  We also construct a naive baseline, referred to as 2D-Lifting, which uses the same inputs as our method but performs editing on 2D images with Lightning-Drag \cite{shi2024instadrag}, followed by reconstruction \cite{kerbl20233dgaussian}.

It should be noted that this is the first work to perform 3DGS drag-based editing in real scenes, and therefore there are no directly comparable baselines. We compare our method with other 3DGS-based text-driven editing approaches, including GS-Editor \cite{chen2024gaussianeditor} and GS-Ctrl \cite{wu2025gaussctrl}, as well as the anchor point-based dragging method SC-GS \cite{huang2024scgs}. Additionally, we construct a naive baseline, 2D-Lifting, which uses the same inputs as our method but performs drag-based editting on 2D images with our guidance model, Lightning-Drag \cite{shi2024instadrag}, followed by 3D reconstruction \cite{kerbl20233dgaussian}.

\textit{\textbf{Datasets.} }
% 为了证明我们方法的编辑能力与场景泛华能力，我们在三个数据集（XXXXXX）上精心选择了包含室内、室外、物体、人脸等五个场景，进行了超过20种编辑方案。
To comprehensively evaluate our method, we select six representative scenes from two datasets:  Mip-NeRF360 \cite{barron2022mipnerf360} and Instruct-NeRF2NeRF \cite{haque2023instruct}. We perform over 20 types of editing tasks on these scenes, which include human faces, indoor objects, and complex outdoor scenes.

% We carefully selected  including indoor, outdoor, objects, and faces, 
% human faces, and complex outdoor scenes
% from three datasets (XXXXXX) and conducted over 20 types of editing tasks.

% \noindent\textbf{Evaluation Metrics.}
% % 美学指标，执行了73个人的评估，gpt-score
% % 由于目前没有广泛认可的三维场景编辑评测数据集，我们执行了User-study，同时为了保证客观性，采用GPT Evaluation Score，用这两种指标来评估编辑结果与质量。使用美学指标来评估编辑后的场景质量。
% As there is currently no widely accepted benchmark dataset for 3D scene editing, we conducted user studies to evaluate the editing results and quality. To ensure objectivity, we also adopted the GPT Evaluation Score \cite{he2023t3benchmark} as an additional metric. Furthermore, we utilized aesthetic evaluation metrics \cite{aa} to assess the quality of the edited scenes.

\subsection{Qualitative Evaluation}
% 我们在图2展示了我们的方法和其他方法： SC-GS \cite{huang2024scgs}, GS-Editor \cite{chen2024gaussianeditor} and GS-Ctrl \cite{wu2025gaussctrl}, 2D-Lifting在三维场景编辑方面的结果。
% 由于目前没有其他方法可以做到三维拖拽，因此为了保证对比的公平性，我们用最大努力对SC-GS、Gaussian Editor、and GS-Ctrl提供引导，帮助他们实现相同的编辑结果。
% 第一列是原三维场景的两个渲染视角，将3D编辑点投影到二维平面进行可视化展示。
% 第一排我们展示了Deformation的拖拽场景: 将脸拽向一侧，可以看到SC-GS会出现场景的撕裂现象，而Gaussian Editor仅仅颜色变深了一点，GaussianCtrl变化更明显一点，但是与Gaussian Editor一样，在几何编辑方面失败了。而2D-Ext虽然可以看到转过头，但是由于不一致现象很严重，造成场景模糊。相比而言，我们的方案成功的将头转向了一侧，同时保持多视角一致性与良好的视觉质量。

Since no other existing methods currently support 3D drag-based editing for real scenes, we make our best effort to guide SC-GS, GS-Editor, and GS-Ctrl to achieve comparable editing results, ensuring a fair comparison. 

Fig. \ref{fig-comparison} presents the results of our method compared to other approaches.
% 编辑结果证明了我们的方案能够实现灵活的编辑，同时保证极高的视觉质量与多视角一致性
The editing results illustrate that our approach enables flexible edits while ensuring high visual quality and multi-view consistency. 
For instance, in the first row, we demonstrate a deformation scenario: dragging the face to one side, which is a highly challenging task.
This is because the operation not only involves rotating the entire head but also requires the facial features, such as the eyes, to change harmoniously and synchronously to maintain consistency and realism.
% 这是因为这个操作不仅仅是头部整体的转动，还需要五官保持和谐的同步变化（比如眼睛）
It can be observed that SC-GS may exhibit unnatural distortions, as well as blurring or tearing of the background caused by dragging. GS-Editor only shows minor darkening in color without effective geometric changes. GS-Ctrl achieves slightly more noticeable changes, but similar to GS-Editor, it fails to perform meaningful geometric editing. 2D-Lifting manages to turn the head, but severe inconsistencies across views result in significant scene blurriness. By contrast, our method successfully turns the head to one side while maintaining better details.
    
    % Fig. \ref{fig-moreresult} 展示了我们的方法在更多场景上的编辑结果，证明我们的方法能够应对复杂场景，并能充分理解拖拽目的并生成生成合理的，高质量的三维拖拽结果。
    % Fig. \ref{fig-moreresult} presents more dragging results of our method on additional scenes, demonstrating its capability to handle complex scenarios. It shows that our method can fully comprehend the users' points prompt and produce plausible, high-quality 3D drag-based outcomes.
    Fig. \ref{fig-moreresult} showcases more results on additional scenes, demonstrating its ability to handle complex scenarios. 
    These results illustrate that our method effectively interprets the control-point prompts and produces plausible, high-quality 3D drag-based edits.

\begin{figure*}[!ht]
    \centering
    \includegraphics[width=\linewidth]{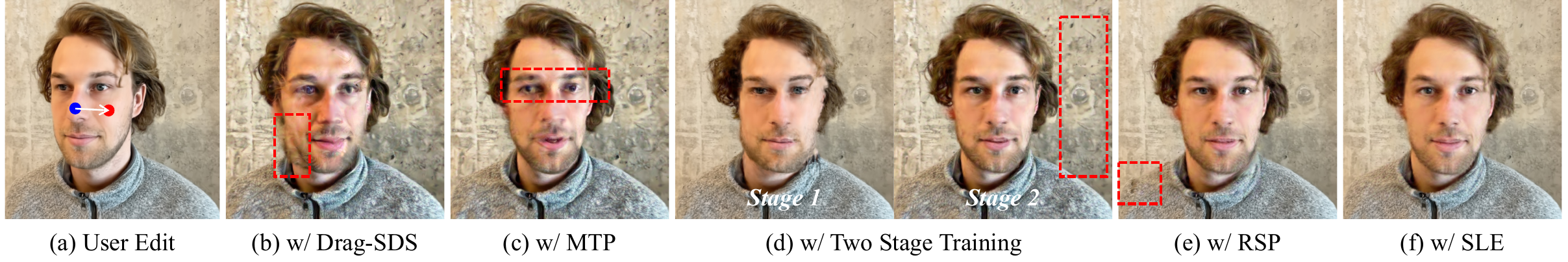}
    \caption{Ablation Study on different modules of DYG. 
    From left to right, new modules are progressively added on top of the previous setup.
    }
    \label{fig-ablation}
\end{figure*}

\begin{figure}
    \centering
    \includegraphics[width=\linewidth]{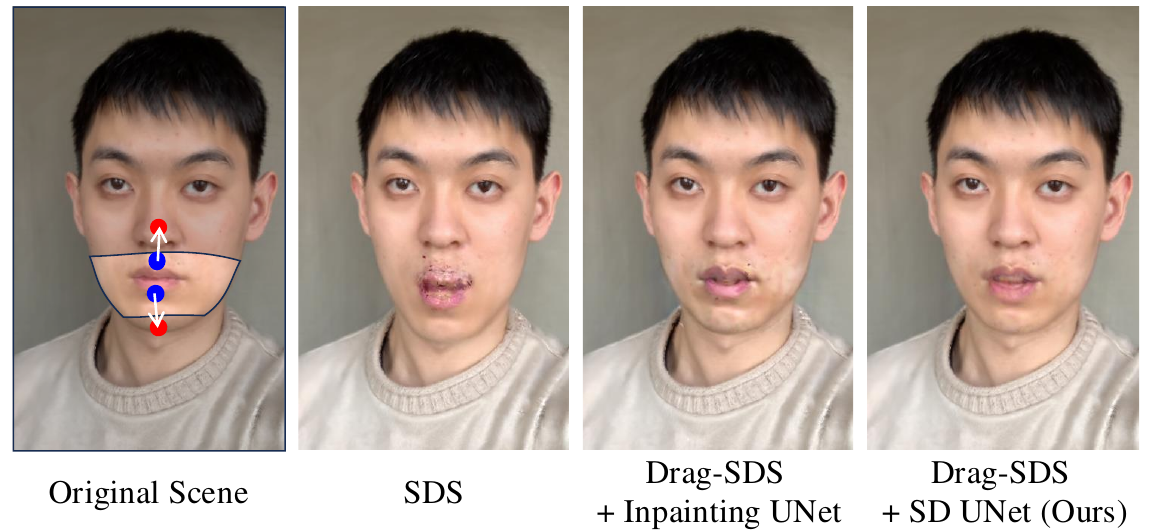}
    \caption{
    Ablation Study of different score distillation loss functions.
    To visualize the 3D mask, we render it as a 2D mask, with the bright region indicating masked area and the darker region unmasked. SDS \cite{poole2022dreamfusion} often causes blurriness in the desired editing area, while Drag-SDS with the inpainting UNet over-focuses on the mask, creating disharmonious color layers. By contrast, our method delivers harmonious editing results.
    }
    \label{fig-sds-ablation}
\end{figure}

\begin{figure}
    \centering
    \includegraphics[width=0.95\linewidth]{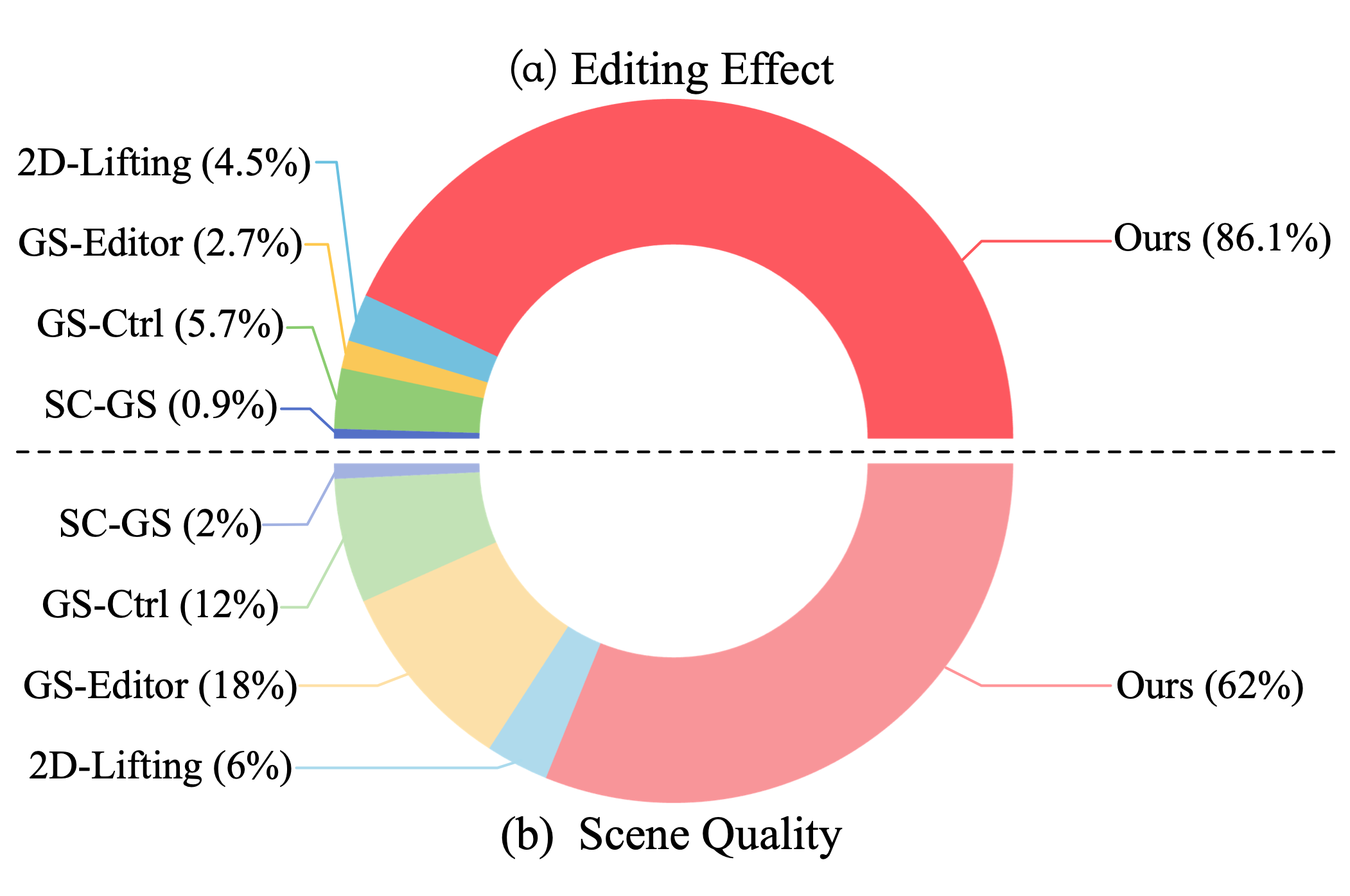}
    \caption{User study of different methods for 3D scene editing.}
    \label{fig-user-study}
\end{figure}

\subsection{Quantitative Evaluation}
As there is currently no widely accepted benchmark dataset for 3D scene editing, we conducted user studies to evaluate the editing results and quality. To ensure objectivity, we also adopted the GPT Evaluation Score as an additional metric. Furthermore, we utilized aesthetic evaluation metrics \cite{schuhmann2022laion} to assess the quality of the edited scenes.

\textit{\textbf{User Study.}} 
We collected survey responses from 75 users, with each questionnaire containing comparisons of 10 edited scenes. Users were asked to select their preferred editing results based on two criteria: Edit Effect and Scene Quality, respectively, resulting in a total of 1,500 votes.
Fig. \ref{fig-user-study} visualizes the results of the user study, showing that 86.1\% and 62\% of users favored our editing results, significantly outperforming other compared methods.

\textit{\textbf{GPT Score.}} 
We utilized GPT-4o to evaluate the editing results of different methods, asking it to rate the results based on three criteria: Scene Quality (SQ), which assesses the visual quality of the edited scene; Editing Effect (EE), which examines whether the editing result meets the intended requirements; and Retention of Initial Features (RIF), which evaluates whether non-edited regions remain unchanged. Scores were assigned on a scale from 0 to 5 for each criterion. 
% 表1中展示了对所有编辑场景的均值得分，按照0.3*SQ + 0.4*EE + 0.3*RIF计算综合分数。 第一行是初始未编辑场景的打分为满分5分。
Table \ref{table-gpt} presents the average scores across all editing scenarios, with the GPT-Overall (GPTO) score calculated as  $0.3 \times \text{SQ} + 0.4 \times \text{EE} + 0.3 \times \text{RIF} $. The first row shows the scores for the initial, unedited scenes, which receive a rating of 5 in all categories. 
% SC-GS由于发生大范围的撕裂现象因此得分很低；GSeditor由于几何变化较小，因此RIF较高但是EE较小；GSCtrl则倾向于让场景颜色变得过于鲜艳，几何编辑则通常失败，因此EE较小；2D-lifting是2D编辑后的重建结果，因此几何编辑结果往往会成功，但是由于多视角不一致，因此场景结果比较模糊，导致SQ较低；我们的方案则在四个指标上都取得了最优的结果。

\begin{table}[!ht]
    \centering
    \caption{\textbf{Evaluation metrics.} 
        % 灰色字体表示初始场景的评估指标
        We report the Scene Quality (SQ), Editing Effect (EE), Retention of Initial Features (RIF), GPT-Overall (GPTO), and Aesthetic (AES) scores for different methods across various scenes. Gray text represents the evaluation metrics of the initial scene.
    }
    \begin{tabular}{ccccc |c}
    \toprule
        ~ & SQ\textuparrow & EE\textuparrow & RIF\textuparrow & GPTO\textuparrow & AES\textuparrow \\ 
    \midrule
        \textcolor{gray}{{Init}} & \textcolor{gray}{5} & ~ & \textcolor{gray}{5} & ~ & \textcolor{gray}{5.53}\\ 
        SC-GS \cite{huang2024scgs} & 2.69 & 2.318 & 2.28 & 2.4182 & 4.14 \\ 
        GS-Editor \cite{chen2024gaussianeditor} & 4.418 & 2.354 & 4.624 & 3.6542 & 5.28  \\ 
        GS-Ctrl \cite{wu2025gaussctrl} & 4.162 & 2.16 & 4.392 & 3.4302 & \textbf{5.38} \\ 
        2D-Lifting & 3.116 & 3.234 & 3.212 & 3.192 & 4.85 \\ 
        Ours & \textbf{4.434} & \textbf{4.42} & \textbf{4.626} & \textbf{4.486} & \underline{5.36} \\ 
    \bottomrule
    \end{tabular}
    \label{table-gpt}
\end{table}

Analyzing in conjunction with Fig. \ref{fig-comparison} and Table \ref{table-gpt},
GS-Editor achieves a relatively high RIF score because of minimal geometric changes but has a low EE score as the edits are less effective. GS-Ctrl often fails in geometric editing, resulting in a low EE score. 2D-Lifting, which reconstructs scenes after 2D dragging, generally succeeds in geometric edits but suffers from multi-view inconsistencies, leading to blurry results and consequently low SQ scores. By contrast, our method achieves the best performance across all four metrics.

 % Scene Quality
 % Editing Effect
 % Retention of Initial Features

\textit{\textbf{Aesthetic Score.}}  
We evaluate the aesthetic quality of 3D edit results using the open-source LAION Aesthetics Predictor, which rates image quality on a 0–10 scale. The rendered images of edited 3D scenes are scored, and the average is reported.
% 如表1最后一列所示，我们的指标在相关方法中取得了良好的结果。值得注意的是，GSCtrl更倾向于将场景整体颜色过饱和化，因此美学指标更高如图4中第一列人脸变白色，因此美学指标更高，但是却没有实现目标的转头编辑操作。
As shown in the last column of Table 1, our method 
% 只比初始美学指标指标降低0.17
decreases by only 0.17 compared to the initial score, achieving better performance compared to SC-GS \cite{huang2024scgs}, GS-Editor \cite{chen2024gaussianeditor}, and 2D-Lifting. Notably, GS-Ctrl tends to oversaturate the overall scene colors, leading to higher aesthetic scores. For example, in the first column of Fig. \ref{fig-comparison}, The face becomes smoother and more cartoon-like, resulting in a higher aesthetic score, but it fails to achieve the intended head-turning editing operation.

\subsection{Ablation Study}

\begin{figure}
    \centering
    \includegraphics[width=\linewidth]{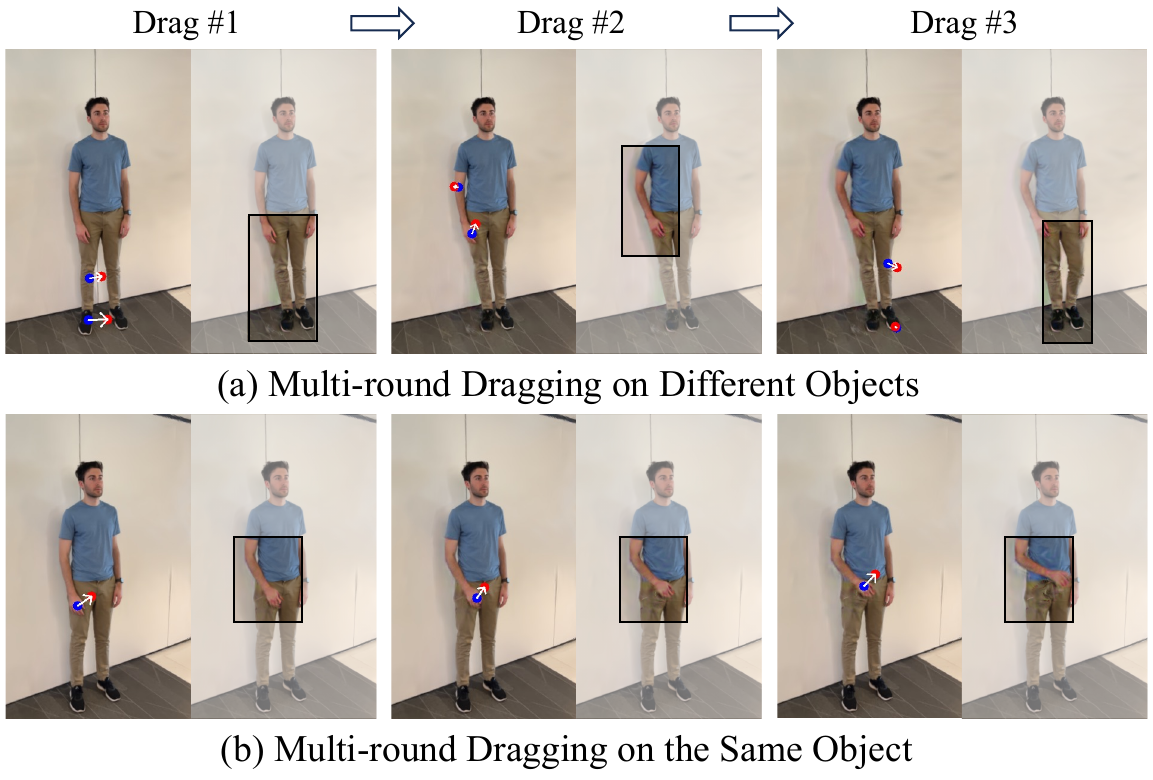}
    \caption{Multi-round Dragging. Each dragging operation is performed based on the results of the previous edit shown on the left.
    % 每组图左侧是场景渲染结果以及drag控制点，右侧是编辑结果。
    % 每组编辑操作都是基于左侧的编辑结果进行拖拽编辑
    }
    \label{fig-multi-round}
\end{figure}

\begin{figure}[t]
    \centering
    \includegraphics[width=\linewidth]{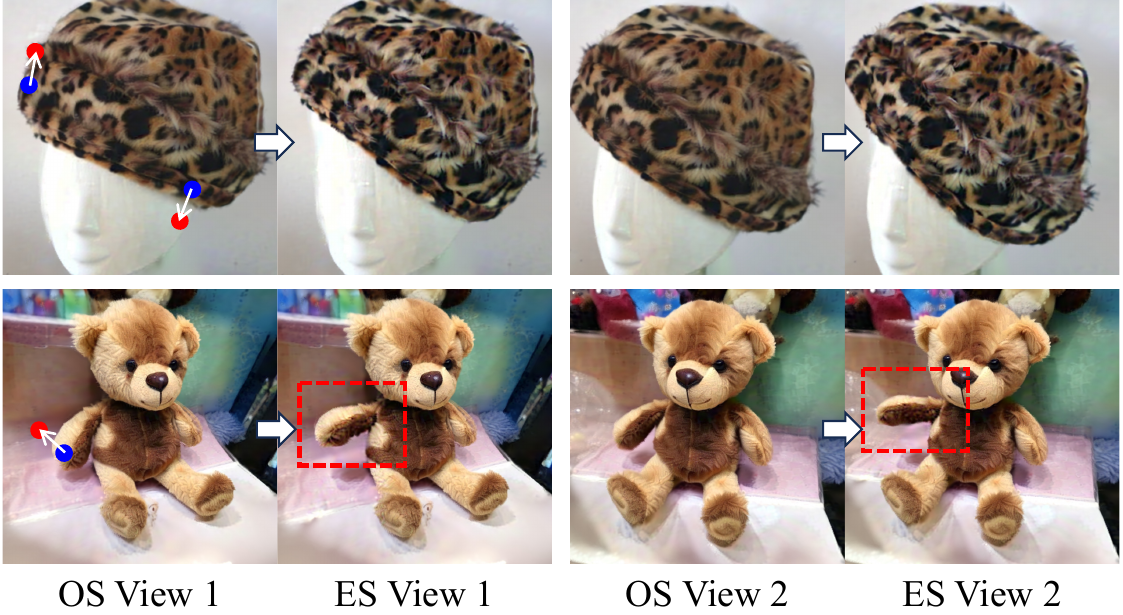}
    % \caption{use Director3D \cite{li2024director3d} to generate two 3DGS scenes with the text prompts “A faux-fur leopard print hat” and “A brown teddy bear in a toy shop,” respectively}
    \caption{Visualization of dragging on the generated scenes. 
    % 我们对每个编辑场景展示原场景和编辑后场景的两个视角。
    For each edit, we show two views of both the original (OS) and edited scenes (ES).
    % 分别是将帽子倾斜和抬起小熊的胳膊
    }
    \label{fig-director3d}
\end{figure}
\paragraph{\textbf{Smooth Geometric Editing module}}
To evaluate the effectiveness of the design of our Smooth Geometric Editing module, we conduct an ablation study and can be shown in Fig. \ref{fig-ablation}.
% 从左到右我们依次在前面的基础上添加新的模块。Fig.\ref{fig-ablation}(b)仅使用Drag-SDS loss function指导优化，此时场景中3DGS所有参数皆可优化。结果显示更倾向于用纹理来拟合目标分布，而不是移动位置，因此人脸左下方会有原本位置的高斯没有完全移动过去，出现伪影；Fig.\ref{fig-ablation}(c)图引入MTPE，同时高斯其他参数可学习。结果显示虽然高斯伪影现象缓解，但是精细部分拖拽效果难以令人满意；Fig.\ref{fig-ablation}（d）引入两阶段优化策略，第一阶段只优化MTPE和RPD，使用高斯的位移来拟合目标分布，搭建编辑结果的脚手架，然后二阶段进行高斯参数的学习，优化场景表示。但是这种方案会造成背景也会被无意中修改，如颜色加深、过饱和等现象。Fig.\ref{fig-ablation}(e)引入了Local Edit策略，但是会出现3Dmask区域和其他区域的裂缝问题。Fig.\ref{fig-ablation}（e）最后我们引入Smooth Local Edit, 达到最高的视觉质量。
From left to right, we progressively add new modules on top of the previous setup.
Fig. \ref{fig-ablation}(b): Only the Drag-SDS loss function is used to guide optimization, with all 3DGS parameters trainable. The results show a tendency to fit the target distribution through texture adjustments rather than position shifts, leaving artifacts where original Gaussians fail to move completely to the target region, such as the lower left corner of the face.
Fig. \ref{fig-ablation}(c): Introducing MTPE with other Gaussian parameters trainable. While this alleviates the Gaussian artifact issue, the fine-grained dragging performance remains unsatisfactory.
Fig. \ref{fig-ablation}(d): A two-stage optimization strategy is introduced. In the first stage, only MTPE and RPD are optimized, aligning the target distribution via Gaussian displacements to build a scaffold for the edited scene. In the second stage, Gaussian parameters are learned to refine the scene representation. However, this approach unintentionally modifies the background, leading to issues like over-saturation and darkened colors.
Fig. \ref{fig-ablation}(e): Local Edit is applied, but it introduces cracks between the 3D mask region and surrounding areas, such as the region around the shoulders.
Fig. \ref{fig-ablation}(f): Finally, the Soft Local Edit strategy is introduced, achieving the highest visual quality with harmonious and consistent results.

% \noindent \textbf{Effectiveness of Drag-SDS.} 
\textit{\textbf{Effectiveness of Drag-SDS.}}
We conduct an ablation study on different distillation loss functions.
All training strategies and modules are identical to the full model, with the only difference being the choice of the distillation loss function.
As shown in Fig. \ref{fig-sds-ablation}, 
% 导致编辑目标区域模糊，无法达到张嘴的动作，相比而言，使用我们提出的Drag-SDS方法能够实现目标的编辑效果，但是由于使用的inpainting方案更倾向于估计mask之内的图像分布，导致出现了区域内外颜色不一致的现象，例如嘴周围颜色更浅，出现了明显地分层。相比而言，我们提出的使用原生SD Unet的方案更关注全局信息，解决了这个问题。
SDS \cite{poole2022dreamfusion} results in a blurred target region, failing to achieve the desired action, such as opening the mouth. In contrast, our proposed Drag-SDS method successfully enables target editing. However, due to its reliance on an inpainting strategy, it focuses more on estimating the image distribution within the mask, leading to inconsistent colors between masked and unmasked regions. For example, the area around the mouth appears lighter, causing noticeable layering artifacts.
In comparison, our approach, which utilizes the original SD \cite{song2020ddim,ho2020ddpm} UNet, pays greater attention to global information and effectively resolves this issue.

\subsection{Multi-round Dragging}
% 由于编辑的复杂性，和多样性，催生了多轮编辑的需求。如图fig-multi-round，我们探索了DYG多轮编辑的可能性，证明了DYG能够can be easily extended to multi-round dragging scenarios，展示了我们的方法具有相当的稳定性。
% 如图1我们在不同的对象（右腿，右胳膊，左腿）上基于前一轮的编辑结果进行后续编辑。
% 如图b我们在同样的对象（逐渐抬起右臂）进行多轮编辑
% , where user can iteratively perform dragging based on the prior output
Due to the complexity and diversity of editing tasks, there arises a demand for multi-round dragging. As shown in Fig. \ref{fig-multi-round}, we explore the potential of extending DYG to multi-round dragging scenarios.
In Fig. \ref{fig-multi-round}(a), we perform sequential edits on different targets (e.g., right leg, right arm, left leg), building upon the results of the previous round. Similarly, Fig. \ref{fig-multi-round}(b) shows multi-round editing applied to the same target (e.g., gradually raising the man's right arm). The results illustrate that DYG can be easily adapted to multi-round dragging scenarios while maintaining notable stability.

\subsection{Dragging for 3D Generative Scenes }
% 除了真实场景，我们也探索了我们的方法在3D生成场景上的编辑结果。如图9，我们用Director3D \cite{s}生成了两个三维高斯场景，分别使用的文本提示是 “A faux-fur leopard print hat”，“A brown teddy bear in a toy shop”。接着我们对他们分别进行了拖拽编辑，我们惊喜的发现，我们的DYG在三维生成结果的编辑上也有良好的泛化能力，能够实现高质量精细拖拽编辑。
In addition to real-world scenes, we also explore the application of our method to editing 3D generative scenes. As shown in Fig. \ref{fig-director3d}, we leverage Director3D \cite{li2024director3d} to generate two scenes with the text prompts “A faux-fur leopard print hat” and “A brown teddy bear in a toy shop,” respectively. Subsequently, we apply drag-based editing to these scenes. To our delight, our DYG  presents strong generalization to 3D generative results, achieving high-quality and precise drag-based editing even in these synthetic scenarios. 
% 更多的生成场景的编辑可以参考Fig. \ref 的最后两排
More examples of scene edits in generated scenarios can be found in the last two rows of Fig. \ref {fig-moreresult}.

\section{Limitation and Conclusion}

\paragraph{\textbf{Limitation}}
%我们的方法从2D基于拖拽的LDM中蒸馏先验知识对3DGS基元属性和位置变形场Triplane进行优化，虽然能够满足多样化的拖拽编辑需求，但编辑结果的视觉质量受限于基于拖拽的LDM的发展。
Our method distills prior knowledge from 2D drag-based LDM to optimize the 3D Gaussian primitives.  Although DYG satisfies a diverse range of 3D drag-based editing requirements, our 3D editing capabilities are inherently limited by the performance of 2D  generative models. Therefore, advancements in 2D generative models can further drive the development of our method.

\textit{\textbf{Conclusion.}}
%我们介绍了DYG，一种基于点的3D高斯场景编辑方法，使得用户能够轻松的编辑3D高斯场景的几何结构。我们通过基于隐式变形场Triplane的两阶段训练策略并利用Drag-SDS对拖拽LDM进行蒸馏，实现了高质量的3D高斯几何编辑。
% We present DYG, an effective drag-based scene editing method that enables users to conveniently edit the  3D Gaussian scenes with 3D masks and control points.
% We present DYG, an effective drag-based scene editing method that enables users to 灵活 细粒度 高质量的
%  edit the  3D Gaussian scenes with 3D masks and control points.
 We present DYG, an effective drag-based scene editing method that enables users to conveniently  perform flexible, fine-grained, high-quality edits on 3D Gaussian scenes using 3D masks and control points. 
 Extensive experiments demonstrate the effectiveness and generalization of our method. 
 % Our future work aims to enabling near-real-time drag-based editing.
% 我们的future work包括提升交互速度，实现近实时的3D drag-based 编辑；另外还可以拓展为4D动态过程，对3D场景编辑得到动态的编辑结果。
 % 大量的实验证明了我们的有效性和泛化性
Our future work includes improving interaction speed to achieve near-real-time 3D drag-based editing. Additionally, it can be extended to 4D dynamic processes, enabling dynamic editing results for 3D scenes.
% 
% 我们的未来工作会提升交互速度，实现近实时的drag-based编辑范式；

% Through a two 
% Through a two-stage training strategy based on the implicit deformation field triplane and the use of our Drag-SDS for distilling the drag-based LDM, we achieve high-quality 3D Gaussian geometry editing.
% 然后写一下未来展望
% \section{Acknowledgments}

% \clearpage

\bibliographystyle{ACM-Reference-Format}
\bibliography{main}


\begin{thebibliography}{46}

%%% the Web address.

\ifx \showCODEN    \undefined \def \showCODEN     #1{\unskip}     \fi
\ifx \showDOI      \undefined \def \showDOI       #1{#1}\fi
\ifx \showISBNx    \undefined \def \showISBNx     #1{\unskip}     \fi
\ifx \showISBNxiii \undefined \def \showISBNxiii  #1{\unskip}     \fi
\ifx \showISSN     \undefined \def \showISSN      #1{\unskip}     \fi
\ifx \showLCCN     \undefined \def \showLCCN      #1{\unskip}     \fi
\ifx \shownote     \undefined \def \shownote      #1{#1}          \fi
\ifx \showarticletitle \undefined \def \showarticletitle #1{#1}   \fi
\ifx \showURL      \undefined \def \showURL       {\relax}        \fi
% invisible to TeX
\providecommand\bibfield[2]{#2}
\providecommand\bibinfo[2]{#2}
\providecommand\natexlab[1]{#1}
\providecommand\showeprint[2][]{arXiv:#2}

\bibitem[Barron et~al\mbox{.}(2022)]%
        {barron2022mipnerf360}
\bibfield{author}{\bibinfo{person}{Jonathan~T. Barron}, \bibinfo{person}{Ben Mildenhall}, \bibinfo{person}{Dor Verbin}, \bibinfo{person}{Pratul~P. Srinivasan}, {and} \bibinfo{person}{Peter Hedman}.} \bibinfo{year}{2022}\natexlab{}.
\newblock \showarticletitle{Mip-NeRF 360: Unbounded Anti-Aliased Neural Radiance Fields}. In \bibinfo{booktitle}{\emph{{IEEE/CVF} Conference on Computer Vision and Pattern Recognition, {CVPR} 2022, New Orleans, LA, USA, June 18-24, 2022}}. \bibinfo{publisher}{{IEEE}}, \bibinfo{pages}{5460--5469}.
\newblock
\urldef\tempurl%
\url{https://doi.org/10.1109/CVPR52688.2022.00539}
\showDOI{\tempurl}


\bibitem[Brooks et~al\mbox{.}(2023)]%
        {brooks2023instructpix2pix}
\bibfield{author}{\bibinfo{person}{Tim Brooks}, \bibinfo{person}{Aleksander Holynski}, {and} \bibinfo{person}{Alexei~A Efros}.} \bibinfo{year}{2023}\natexlab{}.
\newblock \showarticletitle{Instructpix2pix: Learning to follow image editing instructions}. In \bibinfo{booktitle}{\emph{Proceedings of the IEEE/CVF Conference on Computer Vision and Pattern Recognition}}. \bibinfo{pages}{18392--18402}.
\newblock


\bibitem[Chan et~al\mbox{.}(2022)]%
        {chan2022eg3d}
\bibfield{author}{\bibinfo{person}{Eric~R Chan}, \bibinfo{person}{Connor~Z Lin}, \bibinfo{person}{Matthew~A Chan}, \bibinfo{person}{Koki Nagano}, \bibinfo{person}{Boxiao Pan}, \bibinfo{person}{Shalini De~Mello}, \bibinfo{person}{Orazio Gallo}, \bibinfo{person}{Leonidas~J Guibas}, \bibinfo{person}{Jonathan Tremblay}, \bibinfo{person}{Sameh Khamis}, {et~al\mbox{.}}} \bibinfo{year}{2022}\natexlab{}.
\newblock \showarticletitle{Efficient geometry-aware 3d generative adversarial networks}. In \bibinfo{booktitle}{\emph{Proceedings of the IEEE/CVF conference on computer vision and pattern recognition}}. \bibinfo{pages}{16123--16133}.
\newblock


\bibitem[Chen et~al\mbox{.}(2022)]%
        {chen2022tensorf}
\bibfield{author}{\bibinfo{person}{Anpei Chen}, \bibinfo{person}{Zexiang Xu}, \bibinfo{person}{Andreas Geiger}, \bibinfo{person}{Jingyi Yu}, {and} \bibinfo{person}{Hao Su}.} \bibinfo{year}{2022}\natexlab{}.
\newblock \showarticletitle{Tensorf: Tensorial radiance fields}. In \bibinfo{booktitle}{\emph{European conference on computer vision}}. Springer, \bibinfo{pages}{333--350}.
\newblock


\bibitem[Chen et~al\mbox{.}(2025)]%
        {chen2025dge}
\bibfield{author}{\bibinfo{person}{Minghao Chen}, \bibinfo{person}{Iro Laina}, {and} \bibinfo{person}{Andrea Vedaldi}.} \bibinfo{year}{2025}\natexlab{}.
\newblock \showarticletitle{Dge: Direct gaussian 3d editing by consistent multi-view editing}. In \bibinfo{booktitle}{\emph{European Conference on Computer Vision}}. Springer, \bibinfo{pages}{74--92}.
\newblock


\bibitem[Chen et~al\mbox{.}(2024)]%
        {chen2024gaussianeditor}
\bibfield{author}{\bibinfo{person}{Yiwen Chen}, \bibinfo{person}{Zilong Chen}, \bibinfo{person}{Chi Zhang}, \bibinfo{person}{Feng Wang}, \bibinfo{person}{Xiaofeng Yang}, \bibinfo{person}{Yikai Wang}, \bibinfo{person}{Zhongang Cai}, \bibinfo{person}{Lei Yang}, \bibinfo{person}{Huaping Liu}, {and} \bibinfo{person}{Guosheng Lin}.} \bibinfo{year}{2024}\natexlab{}.
\newblock \showarticletitle{Gaussianeditor: Swift and controllable 3d editing with gaussian splatting}. In \bibinfo{booktitle}{\emph{Proceedings of the IEEE/CVF Conference on Computer Vision and Pattern Recognition}}. \bibinfo{pages}{21476--21485}.
\newblock


\bibitem[Fridovich-Keil et~al\mbox{.}(2023)]%
        {fridovich2023kplane}
\bibfield{author}{\bibinfo{person}{Sara Fridovich-Keil}, \bibinfo{person}{Giacomo Meanti}, \bibinfo{person}{Frederik~Rahb{\ae}k Warburg}, \bibinfo{person}{Benjamin Recht}, {and} \bibinfo{person}{Angjoo Kanazawa}.} \bibinfo{year}{2023}\natexlab{}.
\newblock \showarticletitle{K-planes: Explicit radiance fields in space, time, and appearance}. In \bibinfo{booktitle}{\emph{Proceedings of the IEEE/CVF Conference on Computer Vision and Pattern Recognition}}. \bibinfo{pages}{12479--12488}.
\newblock


\bibitem[Guo et~al\mbox{.}(2020)]%
        {guo2020object}
\bibfield{author}{\bibinfo{person}{Michelle Guo}, \bibinfo{person}{Alireza Fathi}, \bibinfo{person}{Jiajun Wu}, {and} \bibinfo{person}{Thomas Funkhouser}.} \bibinfo{year}{2020}\natexlab{}.
\newblock \showarticletitle{Object-centric neural scene rendering}.
\newblock \bibinfo{journal}{\emph{arXiv preprint arXiv:2012.08503}} (\bibinfo{year}{2020}).
\newblock


\bibitem[Haque et~al\mbox{.}(2023)]%
        {haque2023instruct}
\bibfield{author}{\bibinfo{person}{Ayaan Haque}, \bibinfo{person}{Matthew Tancik}, \bibinfo{person}{Alexei~A Efros}, \bibinfo{person}{Aleksander Holynski}, {and} \bibinfo{person}{Angjoo Kanazawa}.} \bibinfo{year}{2023}\natexlab{}.
\newblock \showarticletitle{Instruct-nerf2nerf: Editing 3d scenes with instructions}. In \bibinfo{booktitle}{\emph{Proceedings of the IEEE/CVF International Conference on Computer Vision}}. \bibinfo{pages}{19740--19750}.
\newblock


\bibitem[Hertz et~al\mbox{.}(2022)]%
        {hertz2022prompt}
\bibfield{author}{\bibinfo{person}{Amir Hertz}, \bibinfo{person}{Ron Mokady}, \bibinfo{person}{Jay Tenenbaum}, \bibinfo{person}{Kfir Aberman}, \bibinfo{person}{Yael Pritch}, {and} \bibinfo{person}{Daniel Cohen-Or}.} \bibinfo{year}{2022}\natexlab{}.
\newblock \showarticletitle{Prompt-to-prompt image editing with cross attention control}.
\newblock \bibinfo{journal}{\emph{arXiv preprint arXiv:2208.01626}} (\bibinfo{year}{2022}).
\newblock


\bibitem[Ho et~al\mbox{.}(2020)]%
        {ho2020ddpm}
\bibfield{author}{\bibinfo{person}{Jonathan Ho}, \bibinfo{person}{Ajay Jain}, {and} \bibinfo{person}{Pieter Abbeel}.} \bibinfo{year}{2020}\natexlab{}.
\newblock \showarticletitle{Denoising diffusion probabilistic models}.
\newblock \bibinfo{journal}{\emph{Advances in neural information processing systems}}  \bibinfo{volume}{33} (\bibinfo{year}{2020}), \bibinfo{pages}{6840--6851}.
\newblock


\bibitem[Hu et~al\mbox{.}(2021)]%
        {hu2021lora}
\bibfield{author}{\bibinfo{person}{Edward~J Hu}, \bibinfo{person}{Yelong Shen}, \bibinfo{person}{Phillip Wallis}, \bibinfo{person}{Zeyuan Allen-Zhu}, \bibinfo{person}{Yuanzhi Li}, \bibinfo{person}{Shean Wang}, \bibinfo{person}{Lu Wang}, {and} \bibinfo{person}{Weizhu Chen}.} \bibinfo{year}{2021}\natexlab{}.
\newblock \showarticletitle{Lora: Low-rank adaptation of large language models}.
\newblock \bibinfo{journal}{\emph{arXiv preprint arXiv:2106.09685}} (\bibinfo{year}{2021}).
\newblock


\bibitem[Huang et~al\mbox{.}(2024)]%
        {huang2024scgs}
\bibfield{author}{\bibinfo{person}{Yi-Hua Huang}, \bibinfo{person}{Yang-Tian Sun}, \bibinfo{person}{Ziyi Yang}, \bibinfo{person}{Xiaoyang Lyu}, \bibinfo{person}{Yan-Pei Cao}, {and} \bibinfo{person}{Xiaojuan Qi}.} \bibinfo{year}{2024}\natexlab{}.
\newblock \showarticletitle{Sc-gs: Sparse-controlled gaussian splatting for editable dynamic scenes}. In \bibinfo{booktitle}{\emph{Proceedings of the IEEE/CVF Conference on Computer Vision and Pattern Recognition}}. \bibinfo{pages}{4220--4230}.
\newblock


\bibitem[Kawar et~al\mbox{.}(2023)]%
        {kawar2023imagic}
\bibfield{author}{\bibinfo{person}{Bahjat Kawar}, \bibinfo{person}{Shiran Zada}, \bibinfo{person}{Oran Lang}, \bibinfo{person}{Omer Tov}, \bibinfo{person}{Huiwen Chang}, \bibinfo{person}{Tali Dekel}, \bibinfo{person}{Inbar Mosseri}, {and} \bibinfo{person}{Michal Irani}.} \bibinfo{year}{2023}\natexlab{}.
\newblock \showarticletitle{Imagic: Text-based real image editing with diffusion models}. In \bibinfo{booktitle}{\emph{Proceedings of the IEEE/CVF Conference on Computer Vision and Pattern Recognition}}. \bibinfo{pages}{6007--6017}.
\newblock


\bibitem[Kerbl et~al\mbox{.}(2023)]%
        {kerbl20233dgaussian}
\bibfield{author}{\bibinfo{person}{Bernhard Kerbl}, \bibinfo{person}{Georgios Kopanas}, \bibinfo{person}{Thomas Leimk{\"u}hler}, {and} \bibinfo{person}{George Drettakis}.} \bibinfo{year}{2023}\natexlab{}.
\newblock \showarticletitle{3d gaussian splatting for real-time radiance field rendering}.
\newblock \bibinfo{journal}{\emph{ACM Transactions on Graphics}} \bibinfo{volume}{42}, \bibinfo{number}{4} (\bibinfo{year}{2023}), \bibinfo{pages}{1--14}.
\newblock


\bibitem[Li et~al\mbox{.}(2024)]%
        {li2024director3d}
\bibfield{author}{\bibinfo{person}{Xinyang Li}, \bibinfo{person}{Zhangyu Lai}, \bibinfo{person}{Linning Xu}, \bibinfo{person}{Yansong Qu}, \bibinfo{person}{Liujuan Cao}, \bibinfo{person}{Shengchuan Zhang}, \bibinfo{person}{Bo Dai}, {and} \bibinfo{person}{Rongrong Ji}.} \bibinfo{year}{2024}\natexlab{}.
\newblock \showarticletitle{Director3D: Real-world Camera Trajectory and 3D Scene Generation from Text}.
\newblock \bibinfo{journal}{\emph{arXiv preprint arXiv:2406.17601}} (\bibinfo{year}{2024}).
\newblock


\bibitem[Liu et~al\mbox{.}(2023)]%
        {liu2023zero}
\bibfield{author}{\bibinfo{person}{Ruoshi Liu}, \bibinfo{person}{Rundi Wu}, \bibinfo{person}{Basile Van~Hoorick}, \bibinfo{person}{Pavel Tokmakov}, \bibinfo{person}{Sergey Zakharov}, {and} \bibinfo{person}{Carl Vondrick}.} \bibinfo{year}{2023}\natexlab{}.
\newblock \showarticletitle{Zero-1-to-3: Zero-shot one image to 3d object}. In \bibinfo{booktitle}{\emph{Proceedings of the IEEE/CVF international conference on computer vision}}. \bibinfo{pages}{9298--9309}.
\newblock


\bibitem[Luo et~al\mbox{.}(2024)]%
        {luo2024readout}
\bibfield{author}{\bibinfo{person}{Grace Luo}, \bibinfo{person}{Trevor Darrell}, \bibinfo{person}{Oliver Wang}, \bibinfo{person}{Dan~B Goldman}, {and} \bibinfo{person}{Aleksander Holynski}.} \bibinfo{year}{2024}\natexlab{}.
\newblock \showarticletitle{Readout guidance: Learning control from diffusion features}. In \bibinfo{booktitle}{\emph{Proceedings of the IEEE/CVF Conference on Computer Vision and Pattern Recognition}}. \bibinfo{pages}{8217--8227}.
\newblock


\bibitem[Mildenhall et~al\mbox{.}(2021)]%
        {mildenhall2021nerf}
\bibfield{author}{\bibinfo{person}{Ben Mildenhall}, \bibinfo{person}{Pratul~P Srinivasan}, \bibinfo{person}{Matthew Tancik}, \bibinfo{person}{Jonathan~T Barron}, \bibinfo{person}{Ravi Ramamoorthi}, {and} \bibinfo{person}{Ren Ng}.} \bibinfo{year}{2021}\natexlab{}.
\newblock \showarticletitle{NeRF: Representing Scenes as Neural Radiance Fields for View Synthesis}.
\newblock \bibinfo{journal}{\emph{Commun. ACM}} \bibinfo{volume}{65}, \bibinfo{number}{1} (\bibinfo{year}{2021}), \bibinfo{pages}{99--106}.
\newblock


\bibitem[Mou et~al\mbox{.}(2023)]%
        {mou2023dragondiffusion}
\bibfield{author}{\bibinfo{person}{Chong Mou}, \bibinfo{person}{Xintao Wang}, \bibinfo{person}{Jiechong Song}, \bibinfo{person}{Ying Shan}, {and} \bibinfo{person}{Jian Zhang}.} \bibinfo{year}{2023}\natexlab{}.
\newblock \showarticletitle{Dragondiffusion: Enabling drag-style manipulation on diffusion models}.
\newblock \bibinfo{journal}{\emph{arXiv preprint arXiv:2307.02421}} (\bibinfo{year}{2023}).
\newblock


\bibitem[M{\"u}ller et~al\mbox{.}(2022)]%
        {muller2022instant}
\bibfield{author}{\bibinfo{person}{Thomas M{\"u}ller}, \bibinfo{person}{Alex Evans}, \bibinfo{person}{Christoph Schied}, {and} \bibinfo{person}{Alexander Keller}.} \bibinfo{year}{2022}\natexlab{}.
\newblock \showarticletitle{Instant neural graphics primitives with a multiresolution hash encoding}.
\newblock \bibinfo{journal}{\emph{ACM transactions on graphics (TOG)}} \bibinfo{volume}{41}, \bibinfo{number}{4} (\bibinfo{year}{2022}), \bibinfo{pages}{1--15}.
\newblock


\bibitem[Nie et~al\mbox{.}(2023)]%
        {nie2023blessing}
\bibfield{author}{\bibinfo{person}{Shen Nie}, \bibinfo{person}{Hanzhong~Allan Guo}, \bibinfo{person}{Cheng Lu}, \bibinfo{person}{Yuhao Zhou}, \bibinfo{person}{Chenyu Zheng}, {and} \bibinfo{person}{Chongxuan Li}.} \bibinfo{year}{2023}\natexlab{}.
\newblock \showarticletitle{The blessing of randomness: Sde beats ode in general diffusion-based image editing}.
\newblock \bibinfo{journal}{\emph{arXiv preprint arXiv:2311.01410}} (\bibinfo{year}{2023}).
\newblock


\bibitem[Ost et~al\mbox{.}(2021)]%
        {ost2021neural}
\bibfield{author}{\bibinfo{person}{Julian Ost}, \bibinfo{person}{Fahim Mannan}, \bibinfo{person}{Nils Thuerey}, \bibinfo{person}{Julian Knodt}, {and} \bibinfo{person}{Felix Heide}.} \bibinfo{year}{2021}\natexlab{}.
\newblock \showarticletitle{Neural scene graphs for dynamic scenes}. In \bibinfo{booktitle}{\emph{Proceedings of the IEEE/CVF Conference on Computer Vision and Pattern Recognition}}. \bibinfo{pages}{2856--2865}.
\newblock


\bibitem[Pan et~al\mbox{.}(2023)]%
        {pan2023draggan}
\bibfield{author}{\bibinfo{person}{Xingang Pan}, \bibinfo{person}{Ayush Tewari}, \bibinfo{person}{Thomas Leimk{\"u}hler}, \bibinfo{person}{Lingjie Liu}, \bibinfo{person}{Abhimitra Meka}, {and} \bibinfo{person}{Christian Theobalt}.} \bibinfo{year}{2023}\natexlab{}.
\newblock \showarticletitle{Drag your gan: Interactive point-based manipulation on the generative image manifold}. In \bibinfo{booktitle}{\emph{ACM SIGGRAPH 2023 Conference Proceedings}}. \bibinfo{pages}{1--11}.
\newblock


\bibitem[Park et~al\mbox{.}(2023)]%
        {NEURIPS2023_4bfcebed}
\bibfield{author}{\bibinfo{person}{Yong-Hyun Park}, \bibinfo{person}{Mingi Kwon}, \bibinfo{person}{Jaewoong Choi}, \bibinfo{person}{Junghyo Jo}, {and} \bibinfo{person}{Youngjung Uh}.} \bibinfo{year}{2023}\natexlab{}.
\newblock \showarticletitle{Understanding the Latent Space of Diffusion Models through the Lens of Riemannian Geometry}. In \bibinfo{booktitle}{\emph{Advances in Neural Information Processing Systems}}, \bibfield{editor}{\bibinfo{person}{A.~Oh}, \bibinfo{person}{T.~Naumann}, \bibinfo{person}{A.~Globerson}, \bibinfo{person}{K.~Saenko}, \bibinfo{person}{M.~Hardt}, {and} \bibinfo{person}{S.~Levine}} (Eds.), Vol.~\bibinfo{volume}{36}. \bibinfo{publisher}{Curran Associates, Inc.}, \bibinfo{pages}{24129--24142}.
\newblock
\urldef\tempurl%
\url{https://proceedings.neurips.cc/paper_files/paper/2023/file/4bfcebedf7a2967c410b64670f27f904-Paper-Conference.pdf}
\showURL{%
\tempurl}


\bibitem[Poole et~al\mbox{.}(2022)]%
        {poole2022dreamfusion}
\bibfield{author}{\bibinfo{person}{Ben Poole}, \bibinfo{person}{Ajay Jain}, \bibinfo{person}{Jonathan~T Barron}, {and} \bibinfo{person}{Ben Mildenhall}.} \bibinfo{year}{2022}\natexlab{}.
\newblock \showarticletitle{Dreamfusion: Text-to-3d using 2d diffusion}.
\newblock \bibinfo{journal}{\emph{arXiv preprint arXiv:2209.14988}} (\bibinfo{year}{2022}).
\newblock


\bibitem[Qu et~al\mbox{.}(2024)]%
        {qu2024goi}
\bibfield{author}{\bibinfo{person}{Yansong Qu}, \bibinfo{person}{Shaohui Dai}, \bibinfo{person}{Xinyang Li}, \bibinfo{person}{Jianghang Lin}, \bibinfo{person}{Liujuan Cao}, \bibinfo{person}{Shengchuan Zhang}, {and} \bibinfo{person}{Rongrong Ji}.} \bibinfo{year}{2024}\natexlab{}.
\newblock \showarticletitle{Goi: Find 3d gaussians of interest with an optimizable open-vocabulary semantic-space hyperplane}. In \bibinfo{booktitle}{\emph{Proceedings of the 32nd ACM International Conference on Multimedia}}. \bibinfo{pages}{5328--5337}.
\newblock


\bibitem[Qu et~al\mbox{.}(2023)]%
        {qu2023sgnerf}
\bibfield{author}{\bibinfo{person}{Yansong Qu}, \bibinfo{person}{Yuze Wang}, {and} \bibinfo{person}{Yue Qi}.} \bibinfo{year}{2023}\natexlab{}.
\newblock \showarticletitle{Sg-nerf: Semantic-guided point-based neural radiance fields}. In \bibinfo{booktitle}{\emph{2023 IEEE International Conference on Multimedia and Expo (ICME)}}. IEEE, \bibinfo{pages}{570--575}.
\newblock


\bibitem[Ramesh et~al\mbox{.}(2022)]%
        {ramesh2022hierarchical}
\bibfield{author}{\bibinfo{person}{Aditya Ramesh}, \bibinfo{person}{Prafulla Dhariwal}, \bibinfo{person}{Alex Nichol}, \bibinfo{person}{Casey Chu}, {and} \bibinfo{person}{Mark Chen}.} \bibinfo{year}{2022}\natexlab{}.
\newblock \showarticletitle{Hierarchical text-conditional image generation with clip latents}.
\newblock \bibinfo{journal}{\emph{arXiv preprint arXiv:2204.06125}} \bibinfo{volume}{1}, \bibinfo{number}{2} (\bibinfo{year}{2022}), \bibinfo{pages}{3}.
\newblock


\bibitem[Schuhmann et~al\mbox{.}(2022)]%
        {schuhmann2022laion}
\bibfield{author}{\bibinfo{person}{Christoph Schuhmann}, \bibinfo{person}{Romain Beaumont}, \bibinfo{person}{Richard Vencu}, \bibinfo{person}{Cade Gordon}, \bibinfo{person}{Ross Wightman}, \bibinfo{person}{Mehdi Cherti}, \bibinfo{person}{Theo Coombes}, \bibinfo{person}{Aarush Katta}, \bibinfo{person}{Clayton Mullis}, \bibinfo{person}{Mitchell Wortsman}, {et~al\mbox{.}}} \bibinfo{year}{2022}\natexlab{}.
\newblock \showarticletitle{Laion-5b: An open large-scale dataset for training next generation image-text models}.
\newblock \bibinfo{journal}{\emph{Advances in Neural Information Processing Systems}}  \bibinfo{volume}{35} (\bibinfo{year}{2022}), \bibinfo{pages}{25278--25294}.
\newblock


\bibitem[Shi et~al\mbox{.}(2024a)]%
        {shi2024instadrag}
\bibfield{author}{\bibinfo{person}{Yujun Shi}, \bibinfo{person}{Jun~Hao Liew}, \bibinfo{person}{Hanshu Yan}, \bibinfo{person}{Vincent~YF Tan}, {and} \bibinfo{person}{Jiashi Feng}.} \bibinfo{year}{2024}\natexlab{a}.
\newblock \showarticletitle{InstaDrag: Lightning Fast and Accurate Drag-based Image Editing Emerging from Videos}.
\newblock \bibinfo{journal}{\emph{arXiv preprint arXiv:2405.13722}} (\bibinfo{year}{2024}).
\newblock


\bibitem[Shi et~al\mbox{.}(2024b)]%
        {shi2024dragdiffusion}
\bibfield{author}{\bibinfo{person}{Yujun Shi}, \bibinfo{person}{Chuhui Xue}, \bibinfo{person}{Jun~Hao Liew}, \bibinfo{person}{Jiachun Pan}, \bibinfo{person}{Hanshu Yan}, \bibinfo{person}{Wenqing Zhang}, \bibinfo{person}{Vincent~YF Tan}, {and} \bibinfo{person}{Song Bai}.} \bibinfo{year}{2024}\natexlab{b}.
\newblock \showarticletitle{Dragdiffusion: Harnessing diffusion models for interactive point-based image editing}. In \bibinfo{booktitle}{\emph{Proceedings of the IEEE/CVF Conference on Computer Vision and Pattern Recognition}}. \bibinfo{pages}{8839--8849}.
\newblock


\bibitem[Song et~al\mbox{.}(2020)]%
        {song2020ddim}
\bibfield{author}{\bibinfo{person}{Jiaming Song}, \bibinfo{person}{Chenlin Meng}, {and} \bibinfo{person}{Stefano Ermon}.} \bibinfo{year}{2020}\natexlab{}.
\newblock \showarticletitle{Denoising diffusion implicit models}.
\newblock \bibinfo{journal}{\emph{arXiv preprint arXiv:2010.02502}} (\bibinfo{year}{2020}).
\newblock


\bibitem[Wang et~al\mbox{.}(2024a)]%
        {wang2024gaussianeditor_water}
\bibfield{author}{\bibinfo{person}{Junjie Wang}, \bibinfo{person}{Jiemin Fang}, \bibinfo{person}{Xiaopeng Zhang}, \bibinfo{person}{Lingxi Xie}, {and} \bibinfo{person}{Qi Tian}.} \bibinfo{year}{2024}\natexlab{a}.
\newblock \showarticletitle{Gaussianeditor: Editing 3d gaussians delicately with text instructions}. In \bibinfo{booktitle}{\emph{Proceedings of the IEEE/CVF Conference on Computer Vision and Pattern Recognition}}. \bibinfo{pages}{20902--20911}.
\newblock


\bibitem[Wang et~al\mbox{.}(2023)]%
        {wang2023ripnerf}
\bibfield{author}{\bibinfo{person}{Yuze Wang}, \bibinfo{person}{Junyi Wang}, \bibinfo{person}{Yansong Qu}, {and} \bibinfo{person}{Yue Qi}.} \bibinfo{year}{2023}\natexlab{}.
\newblock \showarticletitle{Rip-nerf: learning rotation-invariant point-based neural radiance field for fine-grained editing and compositing}. In \bibinfo{booktitle}{\emph{Proceedings of the 2023 ACM International Conference on Multimedia Retrieval}}. \bibinfo{pages}{125--134}.
\newblock


\bibitem[Wang et~al\mbox{.}(2024b)]%
        {wang2024prolificdreamer}
\bibfield{author}{\bibinfo{person}{Zhengyi Wang}, \bibinfo{person}{Cheng Lu}, \bibinfo{person}{Yikai Wang}, \bibinfo{person}{Fan Bao}, \bibinfo{person}{Chongxuan Li}, \bibinfo{person}{Hang Su}, {and} \bibinfo{person}{Jun Zhu}.} \bibinfo{year}{2024}\natexlab{b}.
\newblock \showarticletitle{Prolificdreamer: High-fidelity and diverse text-to-3d generation with variational score distillation}.
\newblock \bibinfo{journal}{\emph{Advances in Neural Information Processing Systems}}  \bibinfo{volume}{36} (\bibinfo{year}{2024}).
\newblock


\bibitem[Wu et~al\mbox{.}(2025)]%
        {wu2025gaussctrl}
\bibfield{author}{\bibinfo{person}{Jing Wu}, \bibinfo{person}{Jia-Wang Bian}, \bibinfo{person}{Xinghui Li}, \bibinfo{person}{Guangrun Wang}, \bibinfo{person}{Ian Reid}, \bibinfo{person}{Philip Torr}, {and} \bibinfo{person}{Victor~Adrian Prisacariu}.} \bibinfo{year}{2025}\natexlab{}.
\newblock \showarticletitle{Gaussctrl: Multi-view consistent text-driven 3d gaussian splatting editing}. In \bibinfo{booktitle}{\emph{European Conference on Computer Vision}}. Springer, \bibinfo{pages}{55--71}.
\newblock


\bibitem[Xu and Harada(2022)]%
        {xu2022deforming}
\bibfield{author}{\bibinfo{person}{Tianhan Xu} {and} \bibinfo{person}{Tatsuya Harada}.} \bibinfo{year}{2022}\natexlab{}.
\newblock \showarticletitle{Deforming radiance fields with cages}. In \bibinfo{booktitle}{\emph{European Conference on Computer Vision}}. Springer, \bibinfo{pages}{159--175}.
\newblock


\bibitem[Yang et~al\mbox{.}(2022)]%
        {yang2022neumesh}
\bibfield{author}{\bibinfo{person}{Bangbang Yang}, \bibinfo{person}{Chong Bao}, \bibinfo{person}{Junyi Zeng}, \bibinfo{person}{Hujun Bao}, \bibinfo{person}{Yinda Zhang}, \bibinfo{person}{Zhaopeng Cui}, {and} \bibinfo{person}{Guofeng Zhang}.} \bibinfo{year}{2022}\natexlab{}.
\newblock \showarticletitle{Neumesh: Learning disentangled neural mesh-based implicit field for geometry and texture editing}. In \bibinfo{booktitle}{\emph{European Conference on Computer Vision}}. Springer, \bibinfo{pages}{597--614}.
\newblock


\bibitem[Yang et~al\mbox{.}(2021)]%
        {yang2021learning}
\bibfield{author}{\bibinfo{person}{Bangbang Yang}, \bibinfo{person}{Yinda Zhang}, \bibinfo{person}{Yinghao Xu}, \bibinfo{person}{Yijin Li}, \bibinfo{person}{Han Zhou}, \bibinfo{person}{Hujun Bao}, \bibinfo{person}{Guofeng Zhang}, {and} \bibinfo{person}{Zhaopeng Cui}.} \bibinfo{year}{2021}\natexlab{}.
\newblock \showarticletitle{Learning object-compositional neural radiance field for editable scene rendering}. In \bibinfo{booktitle}{\emph{Proceedings of the IEEE/CVF International Conference on Computer Vision}}. \bibinfo{pages}{13779--13788}.
\newblock


\bibitem[Yang et~al\mbox{.}(2023)]%
        {yang2023lods}
\bibfield{author}{\bibinfo{person}{Xiaofeng Yang}, \bibinfo{person}{Yiwen Chen}, \bibinfo{person}{Cheng Chen}, \bibinfo{person}{Chi Zhang}, \bibinfo{person}{Yi Xu}, \bibinfo{person}{Xulei Yang}, \bibinfo{person}{Fayao Liu}, {and} \bibinfo{person}{Guosheng Lin}.} \bibinfo{year}{2023}\natexlab{}.
\newblock \showarticletitle{Learn to optimize denoising scores for 3d generation: A unified and improved diffusion prior on nerf and 3d gaussian splatting}.
\newblock \bibinfo{journal}{\emph{arXiv preprint arXiv:2312.04820}} (\bibinfo{year}{2023}).
\newblock


\bibitem[Yuan et~al\mbox{.}(2022)]%
        {yuan2022nerf}
\bibfield{author}{\bibinfo{person}{Yu-Jie Yuan}, \bibinfo{person}{Yang-Tian Sun}, \bibinfo{person}{Yu-Kun Lai}, \bibinfo{person}{Yuewen Ma}, \bibinfo{person}{Rongfei Jia}, {and} \bibinfo{person}{Lin Gao}.} \bibinfo{year}{2022}\natexlab{}.
\newblock \showarticletitle{Nerf-editing: geometry editing of neural radiance fields}. In \bibinfo{booktitle}{\emph{Proceedings of the IEEE/CVF Conference on Computer Vision and Pattern Recognition}}. \bibinfo{pages}{18353--18364}.
\newblock


\bibitem[Zhu et~al\mbox{.}(2023)]%
        {zhu2023hifa}
\bibfield{author}{\bibinfo{person}{Junzhe Zhu}, \bibinfo{person}{Peiye Zhuang}, {and} \bibinfo{person}{Sanmi Koyejo}.} \bibinfo{year}{2023}\natexlab{}.
\newblock \showarticletitle{Hifa: High-fidelity text-to-3d generation with advanced diffusion guidance}.
\newblock \bibinfo{journal}{\emph{arXiv preprint arXiv:2305.18766}} (\bibinfo{year}{2023}).
\newblock


\bibitem[Zhuang et~al\mbox{.}(2024)]%
        {zhuang2024tip}
\bibfield{author}{\bibinfo{person}{Jingyu Zhuang}, \bibinfo{person}{Di Kang}, \bibinfo{person}{Yan-Pei Cao}, \bibinfo{person}{Guanbin Li}, \bibinfo{person}{Liang Lin}, {and} \bibinfo{person}{Ying Shan}.} \bibinfo{year}{2024}\natexlab{}.
\newblock \showarticletitle{Tip-editor: An accurate 3d editor following both text-prompts and image-prompts}.
\newblock \bibinfo{journal}{\emph{ACM Transactions on Graphics (TOG)}} \bibinfo{volume}{43}, \bibinfo{number}{4} (\bibinfo{year}{2024}), \bibinfo{pages}{1--12}.
\newblock


\bibitem[Zhuang et~al\mbox{.}(2023)]%
        {zhuang2023dreameditor}
\bibfield{author}{\bibinfo{person}{Jingyu Zhuang}, \bibinfo{person}{Chen Wang}, \bibinfo{person}{Liang Lin}, \bibinfo{person}{Lingjie Liu}, {and} \bibinfo{person}{Guanbin Li}.} \bibinfo{year}{2023}\natexlab{}.
\newblock \showarticletitle{Dreameditor: Text-driven 3d scene editing with neural fields}. In \bibinfo{booktitle}{\emph{SIGGRAPH Asia 2023 Conference Papers}}. \bibinfo{pages}{1--10}.
\newblock


\bibitem[Zou et~al\mbox{.}(2024)]%
        {zou2024triplane}
\bibfield{author}{\bibinfo{person}{Zi-Xin Zou}, \bibinfo{person}{Zhipeng Yu}, \bibinfo{person}{Yuan-Chen Guo}, \bibinfo{person}{Yangguang Li}, \bibinfo{person}{Ding Liang}, \bibinfo{person}{Yan-Pei Cao}, {and} \bibinfo{person}{Song-Hai Zhang}.} \bibinfo{year}{2024}\natexlab{}.
\newblock \showarticletitle{Triplane meets gaussian splatting: Fast and generalizable single-view 3d reconstruction with transformers}. In \bibinfo{booktitle}{\emph{Proceedings of the IEEE/CVF Conference on Computer Vision and Pattern Recognition}}. \bibinfo{pages}{10324--10335}.
\newblock


\end{thebibliography}

% \appendix









\clearpage

\appendix

% \section{Additional Implementational Details}

\section{Implement Details}
\subsection{Training Setup}
For 3DGS reconstruction, we optimized the Gaussians over 7,000 iterations and set the spherical harmonics to degree 0. In our experiment, batch size is set to 4, and the learning rates for Gaussian's color, opacity, scale, and rotation are set to $2.5\times10^{-3}$, $2.5\times10^{-3}$, $2.5\times10^{-4}$, and $2.5\times10^{-3}$, respectively. The shifts in Gaussian position are entirely obtained through the MTP encoder and the RSP decoder, and we set the learning rate for the MTP encoder was set to $1\times10^{-3}$, while the learning rate for RSP was $5\times10^{-4}$. In the first stage, we freeze the Gaussian attributes and only train MTP and RSP. In the second stage, we lower the learning rate of MTP to $1\times10^{-4}$ to stabilize the scene's geometric structure and begin training the other Gaussian attributes. For Drag-SDS, we set $\lambda_{\text{lat}}=1$, $\lambda_{\text{img}}=0.1$, and $\lambda_{\text{lora}}=1$, the learning rate of the learnable embedding $\hat{y}_\emptyset$ set to $1\times10^{-3}$, and the lora rank set to 16 with the learning rate of $5\times10^{-4}$.

\subsection{Two-Stage Dragging}
During the entire training process, we sample the diffusion time-step $t$ using a cosine annealing schedule $t=f(s)=\frac{1}{2}(T_{\text{max}}-T_{\text{min}})(1+\cos(\pi s))+T_{\text{min}}$, where $s$ is the current training epoch ratio, $[T_{\text{max}},T_{\text{min}}]=[0.98,0.02]$ represents the annealing range for the diffusion timestep. Inspired by previous works that utilized diffusion models based on latent variables for image editing\cite{shi2024dragdiffusion,NEURIPS2023_4bfcebed}, we hypothesize that the diffusion model optimize the overall geometric structure of the image at higher diffusion timesteps($>0.7T$), and refining the texture at lower timesteps($<0.7T$). Therefore, we derive the two-stage training epochs based on the diffusion timestep, where the chosen timestep threshold is $T_{\text{threshold}}=0.7$, leading to a first-stage geometric reconstruction training epoch ratio of $s=f^{-1}(T_{\text{threshold}})=0.36$

\subsection{Classifier-Free Guidance}
%我们在这里详细的介绍我们使用的classifier-free guidance(cfg) scale,与通常LDM使用固定的cfg不同，我们遵循cite{shi2024instadrag} 使用cfg退火以避免过饱和问题，具体而言，我们使用cfg反平方退火$\omega(s)=(\omega_{\text{max}}-1)\times(1-s)^2+1$，其中$s$为当前训练轮数比例,$s=\frac{cur\_iteration}{total\_iteration}$,$\omega_{\text{max}}=4$
Here, we provide a detailed description of the classifier-free guidance (CFG) scale we used. Unlike the conventional LDM, which uses a fixed CFG, we follow the approach in \cite{shi2024instadrag} and use CFG annealing to avoid the over-saturation issue. Specifically, we employ a CFG inverse square annealing function:
\begin{equation}
    \omega(s)=(\omega_{\text{max}}-1)\times(1-s)^2+1,
    with \ \omega_{\text{max}}=4
\label{cfg-annealing}
\end{equation}
%我们的目标噪声预测使用cfg，而源噪声预测则不使用cfg
Our target noise prediction uses CFG, while the source noise prediction does not:
\begin{equation}
    \epsilon_{\text{tgt}} = \omega(s)(\epsilon_\theta(z_t, t, y)-\epsilon_\theta(z_t, t, \emptyset))
    +\epsilon_\theta(z_t, t, \emptyset)
\label{cfg-tgt}
\end{equation}

\begin{equation}
    \epsilon_{\text{src}} = \hat{\epsilon}_\phi \left(x_t,t,\hat{y}_\emptyset \right)
\label{cfg-src}
\end{equation}

\subsection{Total Loss}
%我们优化3D高斯的最终损失如下，其中\lambda_\text{rr}=2500
We optimize the total loss for the 3D Gaussian as follows, where $\lambda_\text{d-sds}=1$, $\lambda_\text{rr}=2500$: 
\begin{equation}
    \mathcal{L} = \lambda_\text{drag-sds}\mathcal{L}_{\text{Drag-SDS}}+\lambda_\text{rr}\mathcal{L}_{RR}
    \label{final-loss}
\end{equation}

\begin{figure}
    \centering
    \includegraphics[width=0.8\linewidth]{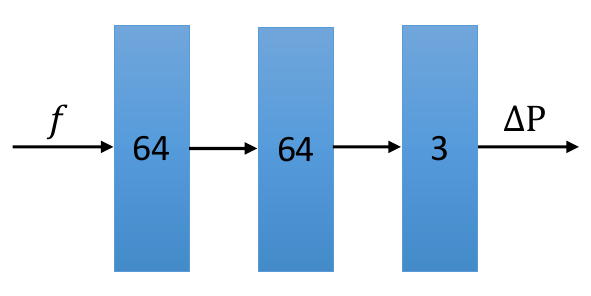}
    \caption{The MLP structure employed in the RSP Decoder.}
    \label{fig-rsp}
\end{figure}

\begin{figure}
    \centering
    \includegraphics[width=\linewidth]{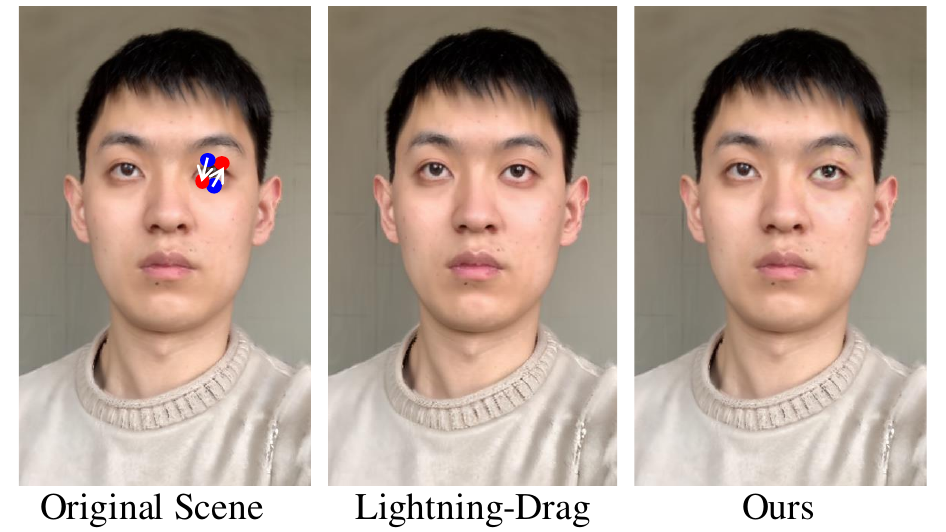}
    %我们的failure case的可视化。我们在对应视角下渲染高斯得到渲染图像和2D mask，并将3D handle points和target points投影到2D作为Lightning-Drag的输入，图中展示了其输出结果。Lighting-Drag在此case中fail，我们的方法以Lightining-Drag作为guidance因此也难以达到期望的结果。
    \caption{Visualization of our failure case. We render the Gaussian in the corresponding viewpoint to obtain the rendered image and 2D mask, and project the 3D handle points and target points into 2D as the input for Lightning-Drag, the figure illustrates the output results. Lightning-Drag fails in this case, and since our method uses Lightning-Drag as guidance, it also struggles to achieve the desired results. }
    \label{fig-fail}
\end{figure}

\section{Details of the RSP Decoder}
%这里我们展示了我们的RSP使用的MLP的结构，MLP接受multi-scale feature $f$作为输入，并以位置变化$\Delta P$作为输出。在我们的RSP Decoder中，有两个这样的MLP,$\mathcal{N}_1$ and $\mathcal{N}_2$,分别用于预测编辑区域和非编辑区域高斯的位置变化。在我们的实验中，我们发现将直接使用$\mathcal{N}_2$预测非编辑区域位置变化会导致严重的几何撕裂问题，为此我们设计了Eq. (\ref{RSP-decoder-eq})防止此问题。RSP所使用的MLP结构
Fig. \ref{fig-rsp} present the structure of the MLP employed in our RSP. The MLP accepts multi-scale features $f$ as input and outputs positional shifts $\Delta P$. In our RSP Decoder, there are two such MLPs, $\mathcal{N}_1$ and $\mathcal{N}_2$, which are used to predict positional shifts of Gaussians in the desired and undesired editing area, respectively. In our experiments, we found that directly using $\mathcal{N}_2$ to predict positional sifts in undesired editing area leads to severe geometric tearing issues. To address this, we designed Eq. (\ref{RSP-decoder-eq}) to prevent such problems.

\begin{equation}
    \Delta p = 
\begin{cases}
\mathcal{N}_1(f), & \text{for } g_i \in \mathcal{G}_{m}, \\
\text{sg}(\mathcal{N}_1(f)) + \mathcal{N}_2(\text{sg}(f)), & \text{for } g_i \in \mathcal{G}_{um},
\end{cases}
\label{RSP-decoder-eq}
\end{equation}

\section{Failure Case}
%因为我们使用了预训练基于拖拽的LDM Lightning-Drag，因此我们的方法也继承了其失败的情况。图中展示了用户输入并期望让他闭上眼睛，我们的方法和Lightning-Drag在此case中均fail。
Since we used the pretrained drag-based LDM Lightning-Drag\cite{shi2024instadrag}, our method also inherits its failure cases. The Fig. \ref{fig-fail} shows the user input where the user expects his eye to be closed, and both Lightning-Drag and our method fail in this case.

\end{document}